\documentclass[12pt, a4paper]{article}

\usepackage{amsmath,amssymb}
\usepackage{jheppub}
\usepackage{etoolbox}
\usepackage{bm}
\usepackage{ulem}
\usepackage{subfig}
\usepackage[utf8]{inputenc}
\usepackage[english]{babel}

\usepackage{axodraw4j}
\usepackage{pstricks}
\usepackage{color}

\newcommand{\p}{\partial}
\newcommand{\Tr}{\text{Tr}}

\def\({\left(} \def\){\right)}
\def\[{\left[} \def\]{\right]}
 
\def\del{{\partial}}

\def\g{\gamma}

\newcommand{\be}{\begin{equation}}
\newcommand{\ee}{\end{equation}}
\newcommand{\bea}{\begin{eqnarray}}
\newcommand{\eea}{\end{eqnarray}}
\newcommand{\ba}{\begin{eqnarray}}
\newcommand{\ea}{\end{eqnarray}}
\newcommand{\nn}{\nonumber \\}

\newcommand{\beq}{\begin{equation}}
\newcommand{\eeq}{\end{equation}}
\newcommand{\beqa}{\begin{eqnarray}}
\newcommand{\eeqa}{\end{eqnarray}}
\newcommand{\beqar}{\begin{eqnarray*}}
\newcommand{\eeqar}{\end{eqnarray*}}

\def\f{\frac}
\def\p{\partial}
\def\nn{\nonumber}
\def\lan{\langle}
\def\ran{\rangle}

\def\l{\lambda}

\def\s{\sigma}

\def\g{\gamma}
\def\G{\Gamma}

\def\ba{\bar{a}}

\def\df{\Delta_{\phi}}
\def\ds{\Delta_{\sigma}}

\def\d{\delta}

\def\mo{\mathcal{O}}

\def\Tr{\text{Tr}}

\def\ms{\Sigma}
\def\hs{\hat{\sigma}}
\def\hp{\hat{\phi}}

\allowdisplaybreaks[3]

\makeatletter
\patchcmd{\maketitle}{\@fpheader}{}{}{}
\makeatother

\begin{document}

\title{Long-Range Vector Models at Large N}

\author[]{Noam Chai,}
\author[]{Mikhail Goykhman,}
\author[]{Ritam Sinha}
\affiliation[]{The Racah Institute of Physics, The Hebrew University of Jerusalem, \\ Jerusalem 91904, Israel}

\abstract{We calculate various CFT data for the $O(N)$ vector model with the long-range interaction, working
at the next-to-leading order in the $1/N$ expansion.
Our results provide additional evidence for the existence of conformal symmetry at the long-range fixed point,
as well as the continuity of the CFT data at the long-range to short-range crossover point $s_\star$ of the
exponent parameter $s$. We also develop the $N>1$ generalization of the recently proposed IR duality
between the long-range and the deformed short-range models, providing further evidence
for its non-perturbative validity in the entire region $d/2<s<s_\star$.}

\emailAdd{noam.chai@mail.huji.ac.il}
\emailAdd{michael.goykhman@mail.huji.ac.il}
\emailAdd{ritam.sinha@mail.huji.ac.il}

\maketitle
\section{Introduction}
\label{sec:Intro}
Long-range spin models provide a simple generalization of the usual Ising spin chain with the nearest neighbor interactions. However, they are known to display a much richer phase structure. The long-range Ising
(LRI) model was introduced by Dyson more than fifty years ago to describe spontaneous symmetry breaking and long-range order in a $1d$ ferromagnetic spin chain \cite{Dyson:1968up}. It features a spin-spin interaction that decays as a power law, controlled by a  positive-valued exponent
 $s$. In a general dimension $d$, the Hamiltonian for the LRI model is given by
\begin{align}
\label{LRI hamiltonian}
  H = -J\sum_{i,j}\f{s_i s_j}{|i-j|^{d+s}}\,,
\end{align}
where $J>0$ governs the strength of interaction and $s_i=\pm1$ are the Ising spins.  This model has been subjected to numerous tests, both analytical and numerical, and was found to exhibit a second-order phase transition in low-dimensional spin chains \cite{Fisher:1972zz,PhysRevLett.37.1577, Aizenman1988}. The corresponding long-range critical regime was found in the window $d/2<s< s_\star$
for the parameter $s$.
For the range $s\leq d/2$, the model is described by a Gaussian mean-field theory (MFT), whereas for $s\geq s_\star$, the model transitions to its short-range counterpart, given by the usual Ising CFT. 

It is easy to identify these phases in the parameter space of $s$ in the continuum description of (\ref{LRI hamiltonian}), obtained using the Landau-Ginzburg technique,
\begin{align}
 S \propto -\int\,d^dx \int\,d^dy\f{\phi(x)\phi(y)}{|x-y|^{d+s}}\,.
 \label{Gaussian}
\end{align}
Using \eqref{Gaussian}, the scaling dimension of the field $\phi$ is found to be $\Delta_\phi=(d-s)/2$.
It follows that we need $s\leq2$, in order for the long-range model to be unitary. 
One can subsequently deform the above action by adding to it a usual local quartic interaction term \cite{Fisher:1972zz},
\begin{align}
 S \propto -\int\,d^dx \int\,d^dy\f{\phi(x)\phi(y)}{|x-y|^{d+s}} + g\,\int\,d^d x\,\phi(x)^4\,.
 \label{phi4}
\end{align}
When $s\leq d/2$, the quartic interaction is irrelevant,
and the model is described by the Gaussian MFT. When $s>d/2$, the quartic interaction becomes relevant, and triggers an RG flow, taking the theory away from the Gaussian fixed point. One can perform a systematic perturbative expansion near $s=d/2$ to identify a non-trivial interacting IR fixed point of this flow. This fixed point has been studied extensively in the literature (see \cite{Brydges:2002wq,Abdesselam:2006qg,Slade:2016yer,mitter,Benedetti:2020rrq} for a rigorous renormalization group derivation of this fixed point), and much is known about the corresponding critical exponents. The scaling dimension of the operator $\phi$
at this long-range fixed point is exact, being protected by the bi-local kinetic term in (\ref{phi4})
from receiving anomalous contributions.
 This fixed point is also characterized by the lack of a local stress-energy tensor. The latter fact makes it rather difficult to ascertain whether the long-range fixed point in fact enjoys the full conformal symmetry.
 Strong support in favor of existence of conformal symmetry at the long-range fixed point has recently been provided in  \cite{Paulos:2015jfa,Behan:2017dwr,Behan:2017emf,Behan:2018hfx}.



As we reviewed above, the long-range critical regime is found at the end of an RG flow
in (\ref{phi4}) for the exponent parameter $s$ taking values in the range $d/2<s<s_\star$. The model crosses over to the short-range CFT regime when $s\geq s_\star$.\footnote{The short-range CFT at $s>s_\star$ is the critical short-range
 Ising model plus
a decoupled sector consisting of a generalized free field \cite{Behan:2017emf,Behan:2018hfx}, as we review below.}
While the dimension of $\phi$ in the long range CFT is $(d-s)/2$, its corresponding dimension in the short-range CFT is $d/2-1+\gamma_{\hat\phi}$, with $\gamma_{\hat\phi}$ being the anomalous dimension of the short-range field $\hat\phi$
(in this paper we will typically put a hat on top of the letter standing for the short-range field,
whenever the same letter without a hat is used to denote the long-range field).
Continuity at $s_\star$ then implies that $s_\star = 2-2\gamma_{\hat\phi}$ \cite{Sak1973,PhysRevB154344}. Notice that since $\gamma_{\hat\phi} > 0$
in the usual short-range CFT, then $s_\star < 2$, implying that interactions tighten the upper bound on $s$.
In fact, it follows that all CFT data, not just the conformal dimension of $\phi$, are continuous across the crossover point $s_\star$ \cite{Sak1973,PhysRevB154344}. In this paper we will illustrate such continuity by carrying out explicit calculations
of CFT data in the $O(N)$ generalization of the model (\ref{phi4}), working at the next-to-leading
order in $1/N$ expansion.\\

We begin by considering an $O(N)$ version of the non-local MFT, and perturb it by a quartic interaction. To access the  strongly-interacting IR regime of this model away from the weakly-coupled behavior near $s=d/2$, we
study it in the $1/N$ expansion. By employing the Hubbard-Stratonovich formalism
we can write down the corresponding action
of the model as
\begin{align}
 S\propto-\int\,d^dx \int\,d^dy\f{\phi_i(x)\phi_i(y)}{|x-y|^{d+s}} + \int\,d^d x\,\left(-\f{1}{4g}\s^2 + \f1{\sqrt{N}}\s\phi^2\right)\,,
 \label{HS-action}
\end{align}
where $\s$ is the Hubbard-Stratonovich field.
The model (\ref{HS-action}) has been studied previously in the literature, where critical exponents such as the scaling dimensions of $\phi$ and $\s$ have been calculated \cite{Brezin_2014,Gubser:2017vgc, Giombi:2019enr}. However, no other CFT data in the $1/N$ expansion is available for the long-range fixed point. One of the goals of this paper is to fill this gap, by calculating some large $N$ CFT data, including anomalous dimensions of composite operators, and  various OPE
coefficients.
We perform most of our calculations at the next-to-leading order in the $1/N$ expansion.
Our results lend a strong support to the statement that the long-range fixed point enjoys the full conformal symmetry, even for the values of $s$ that are beyond the scope of a perturbative regime in the vicinity of $s=d/2$.

Following our calculations of CFT data in the long-range critical $O(N)$ vector model, we proceed to establish
explicitly that the obtained anomalous dimensions and OPE coefficients are continuous at 
 $s=s_\star$. In fact, at $s=s_\star$ the long-range critical vector model crosses over to the
 short-range critical vector model plus a generalized free field $\chi$ \cite{Behan:2017dwr,Behan:2017emf}. The role of the decoupled field $\chi$ becomes important when one matches the complete spectrum across
 the crossover point $s_\star$. As it was pointed out in \cite{Behan:2017dwr,Behan:2017emf},
 some operators  
 in the long-range model, such as $\phi^3$,  exhibit an apparently discontinuous behavior at $s_\star$, having
 no counterpart in the spectrum of the short-range critical vector model. Introducing the generalized free field $\chi$ allows
 one to construct such operators when $s>s_\star$, thereby allowing for the complete match of the spectrum. Note that there has been some debate regarding the smooth transition of CFT data across this long-range--short-range crossover point \cite{Picco:2012ak,Blanchard:2012xv}. However, our calculation provides complementary support to the numerous theoretical calculations and Monte-Carlo simulations which predicted such a smooth transition at the long-range--short-range crossover point  \cite{Honkonen:1988fq, Honkonen:1990mr,Luijten2002BoundaryBL}.\footnote{In this paper, whenever we refer to the long-range--short-range crossover point $s_\star$, we mean the point of
 crossover between the long-range critical vector model, and the short-range critical vector model with the decoupled
 generalized free field.} \\


While the long-range CFT exists in the perturbative regime near the lower bound $s=d/2$ of the long-range
window $d/2<s<s_\star$, making it amenable to the Wilson-Fisher kind of $\epsilon$-expansion
for $s=(d+\epsilon)/2$,
its behavior near the upper bound $s=s_\star$ is strongly-interacting.
It was proposed in \cite{PhysRevB154344} that the long-range CFT near $s=s_\star$ can be accessed by perturbing
the short-range CFT with a bi-local kinetic term for the field $\hat\phi$. Such a kinetic term
crosses over from being irrelevant when $s>s_\star$, to being relevant when $s<s_\star$.
Consequently, as $s$ is lowered past $s_\star$, an RG flow brings the model to a long-range regime in the IR.
While such a bi-local kinetic term appears to manifest a perturbative behavior in the vicinity of $s_\star$,
it is unclear how to carry out this perturbative expansion. This is primarily due to a lack of proper understanding of how to perform conformal perturbation theory involving non-local perturbations.


Recently a new weakly-coupled description near the long-range--short-range crossover point $s_\star$ was proposed in \cite{Behan:2017dwr,Behan:2017emf}.  
In this description the non-local perturbation of \cite{PhysRevB154344} has been traded for a local one, at the cost of introducing a generalized free field $\chi$ with scaling dimension $\Delta_\chi = (d+s)/2$ into the model. The corresponding action is given by \cite{Behan:2017dwr,Behan:2017emf}
\begin{align}
 S = S_{\text{CFT}} + \l\int\,d^dx \,\hat\phi\chi + \int d^dx\int d^dy\, \f{\chi(x)\chi(y)}{|x-y|^{d-s}}\,,
 \label{short-range}
\end{align}
where $\hat\phi$ is the spin field and $S_{\text{CFT}}$ is the action of the original
short-range CFT model, and $\chi$ is a generalized free field coupled to it.
The interacting term $\hat\phi\chi$ with the coupling $\lambda$ has the leading order scaling dimension $d-\delta$, where $\delta = (s_\star-s)/2$. This perturbation is therefore slightly relevant for small positive $\delta$, that is, for $s$ slightly below the long-range--short-range crossover point $s_\star$. The flow triggered by this perturbation ends at a long-range fixed point in the IR, near $s=s_\star$. Although two different UV descriptions \eqref{phi4} and \eqref{short-range} flow to the long-range fixed points in two non-overlapping perturbative regimes, near $s=d/2$ and $s=s_\star$ respectively, it has been argued in \cite{Behan:2017dwr,Behan:2017emf} that these are in fact the same long-range fixed points for all $d/2<s<s_\star$.
This furnishes a non-trivial example of an \textit{IR duality}. As part of matching the spectra of these two models, the new d.o.f. $\chi$ on the deformed short-range side of the duality has been suggested to be dual to the operator $\phi^3$ in the long-range CFT \cite{Behan:2017dwr,Behan:2017emf}.


One can further take advantage of the perturbative behavior of the coupling $\lambda$
near $s=s_\star$ to find
the anomalous dimension of the stress-energy tensor $T_{\mu\nu}$ within conformal perturbation theory. Importantly, a non-trivial anomalous dimension
of $T_{\mu\nu}$ supplies further evidence that the fixed point at the end of an RG flow triggered by the coupling $\lambda$ is indeed a long-range CFT.
Additional evidence for the proposed IR duality is given by matching the ratios of the three-point function
amplitudes (OPE coefficients) $\langle\phi{\cal O}_1{\cal O}_2\rangle$, $\langle\phi^3{\cal O}_1{\cal O}_2\rangle$
in the long-range model, and $\langle\hat\phi{\cal O}_1{\cal O}_2\rangle$, $\langle\chi{\cal O}_1{\cal O}_2\rangle$
in the deformed short-range model, for arbitrary operators ${\cal O}_{1,2}$ \cite{Behan:2017dwr,Behan:2017emf}.\\

Inspired by the construction of \cite{Behan:2017dwr,Behan:2017emf}, we propose that the long-range $O(N)$
critical vector model for $d/2<s<s_\star$ admits a dual description in terms of a deformed
short-range CFT with the action
\begin{align}
 S = a\,\int d^dx\int d^dy\, \f{\chi ^i(x)\chi ^i(y)}{|x-y|^{d-s}}+ \f1{\sqrt{N}}\,\int\,d^dx\,\ms\,\Phi^2+\int\,d^dx\, \Phi ^i\chi ^i  \,,
 \label{dual-action}
\end{align}
where $\Phi^i$, $i=1,\dots, N$ is the spin field, $\Sigma$ is the Hubbard-Stratonovich field, and $\chi^i$, $i=1,\dots, N$
is a generalized free vector field.\footnote{The coefficient $a$ reflects a particular choice of
conventions regarding normalization of $\chi^i$, and will be fixed below.}
We will argue that the models \eqref{dual-action}
and (\ref{HS-action}) are equivalent, up to an identification of the d.o.f.
\begin{align}
\label{intro phi Phi relation}
 \Phi\leftrightarrow \phi, \hspace{1cm}\ms\leftrightarrow \sigma\,.
\end{align}
These two models, therefore, provide dual descriptions of the same long-range IR critical regime, for all values of $d/2<s<s_\star$.

It then follows that the composite operator $\s\phi$ is dual to the
field $\chi$ on the other side of the duality. While this can be seen as a direct
consequence of the relations (\ref{intro phi Phi relation}) and the e.o.m.,
one can additionally carry out a consistency check of the proposed duality,
by matching the amplitudes of the three-point functions involving the operators
$\phi$, $\s\phi$ in the long-range model, and $\hat\phi$, $\chi$ \cite{Behan:2017dwr,Behan:2017emf}.
As we move to $s>s_\star$, beyond the scope of the long-range critical regime, the field $\chi$
decouples into a separate generalized free field sector. Its existence however
guarantees continuity of spectrum across the crossover point $s_\star$ \cite{Behan:2017dwr,Behan:2017emf}.

Following \cite{Behan:2017dwr,Behan:2017emf},
we further use our newly proposed dual IR long-range CFT description to calculate the anomalous dimension and trace of the stress-energy tensor. This calculation additionally
reveals that near $s=s_\star$ our dual long-range CFT is perturbatively in $1/N$ close to the short-range model in the UV,
by relating anomalous dimension of the stress-energy tensor to the anomalous dimension of the spin field $\phi$.


This paper is organized as follows. In section~\ref{sec:setup}, we define the action for the long-range critical $O(N)$ vector model at large $N$, and set up our conventions 
that will subsequently be used throughout the rest of the paper. In section~\ref{sec:two-pt}, we compute the anomalous dimensions and amplitude corrections of the lowest scalar primaries $\phi$ and $\s$, up 
to the next-to-leading order in $1/N$ expansion. In section~\ref{sec:conf-inv}, we begin by calculating the anomalous dimensions of composite scalar primaries $\s^n$, at the next-to-leading order in $1/N$, followed by the
calculation of the cross-correlator  $\langle\s \s^2\rangle$.
We establish that such a cross-correlator in fact vanishes, unlike its short-range counterpart.
In particular, this calculation provides additional evidence that the long-range fixed point enjoys the full conformal symmetry. In section~\ref{sec:OPE coefficients}, we calculate the OPE coefficients $C_{\phi\phi\s}$ and $C_{\s\s\s}$, and the corresponding conformal triangles, working at the next-to-leading order in $1/N$.
In section~\ref{sec: crossover to short range}, we explicitly demonstrate the continuity of all the calculated
 long-range CFT data at the long-range--short-range crossover point $s_\star$.
 In section~\ref{sec: duality}, we study a short-range critical $O(N)$ vector model coupled to a generalized free field $\chi^i$, which we propose to be
  dual to the long-range critical vector model in the IR.
 We discuss our results and outline some future research directions in section~\ref{sec:discussion}.

\section{Set-up}
\label{sec:setup}

In this section we review the basic setup of the long-range 
critical $O(N)$ vector model that we will be studying in this paper.
This model describes dynamics of a multiplet of $N$ scalar fields
$\phi_i$, $i=1,\dots,N$ in the fundamental
representation of the $O(N)$ symmetry group, featuring a self-interaction
with the quartic coupling constant $g$. We will be interested in the interacting critical
regime of this model.
In the free regime, $g=0$, the model sits at its Gaussian fixed point,
where it is described by the MFT, 
with the bi-local kinetic action term analogous to \eqref{Gaussian},
\begin{align}
  S_0=C(s) \int d^d x\,\int d^d y\,\f{\phi_i(x)\phi_i(y)}{|x-y|^{d+s}}\,.
  \label{bi-mean-field}
\end{align}

This class of models is parametrized by the exponent $s$, characterizing
the power-law long-range kinetic term in the action (\ref{bi-mean-field}).
The model (\ref{bi-mean-field}) is defined for $s < 2$, which ensures that
the bi-local term is more relevant than the local kinetic term $\frac{1}{2}\partial\phi_i\partial\phi_i$.
The latter dominates for $s>2$, when the theory crosses over to the local model (free short-range $O(N)$ vector model).
In fact, when $s=2$, the model (\ref{bi-mean-field}) is equivalent to
the short-range $O(N)$ vector model, as it is easiest to see in momentum space.
For $s=1$, the action (\ref{bi-mean-field}) reduces, interestingly, to a boundary CFT (bCFT) with a free scalar field in the bulk,
see \cite{Giombi:2019enr} for some recent developments.

We will use the conventional choice of the factor $C(s)$ such that the propagator of $\phi_i$ in momentum space is
normalized to unity (from now on we skip keeping track of the $O(N)$ indices and the associated Kronecker
symbol whenever it does not cause an ambiguity),
\begin{equation}
\langle \phi(p)\phi(q)\rangle = (2\pi)^d\,\delta^{(d)}(p+q)\,\frac{1}{(p^2)^\frac{s}{2}}\,.
\end{equation}
Using the Fourier transform relation
\begin{equation}
\int\frac{d^dk}{(2\pi)^d}\,e^{ik\cdot x}\,\frac{1}{(k^2)^{\frac{d}{2}-\Delta}}
=\frac{2^{2\Delta-d}}{\pi^\frac{d}{2}}\,\frac{1}{A(\Delta)}\,\frac{1}{|x|^{2\Delta}}\,,
\end{equation}
where we defined
\begin{equation}
A(\Delta) = \frac{\Gamma\left(\frac{d}{2}-\Delta\right)}{\Gamma(\Delta)}\,,
\end{equation}
we obtain\footnote{Notice that naively setting $s=2$ renders a singularity in $C(s)$.
This artifact is lifted in momentum space.}
\begin{align}
\label{C(s)}
 C(s) = \f{2^{s-1}}{\pi^{d/2}}\f{\G\left(\f{d+s}2\right)}{\G\left(-\f{s}2\right)}\,.
\end{align}
The field $\phi$ has the dimension
\begin{equation}
\label{Delta phi}
\Delta_\phi = \frac{d-s}{2}\,,
\end{equation}
and its propagator in coordinate space has the form
\begin{equation}
\langle \phi(x)\phi(0)\rangle = \frac{C_\phi}{|x|^{2\Delta_\phi}}\,,
\end{equation}
where we defined
\begin{equation}
\label{Cphi def}
C_\phi = \frac{1}{2^s\pi^\frac{d}{2}}\,\frac{\Gamma\left(\frac{d-s}{2}\right)}{\Gamma\left(\frac{s}{2}\right)}\,.
\end{equation}

Perturbing the free theory (\ref{bi-mean-field}) by a quartic self-interaction we obtain
an interacting field theory with the action
\begin{align}
\label{starting bi-local interaction action}
 S = C(s) \int d^d x\,\int d^d y\,\f{\phi(x)\phi(y)}{|x-y|^{d+s}}+ \f{g}N\int\,d^d x\, (\phi^2)^2\,.
\end{align}
Setting $g=0$, we recover the free action (\ref{bi-mean-field}).
The quartic coupling constant $g$ is relevant when $s>d/2$, and it triggers a flow to a fixed point in the IR.
This can be established perturbatively by a Wilson-Fisher kind of $\epsilon$-expansion
for $s=d/2+\epsilon$. Together with the non-locality constraint $s<2$, this
restricts us to consider $d<4$. When $s<d/2$, the quartic coupling $g$
is irrelevant, and the theory flows to a fixed point in the UV limit,
albeit the resulting model suffers from instabilities. This situation is analogous 
to the Wilson-Fisher fixed point of the short-range $O(N)$ vector model in
$2<d<4$ and $4<d<6$ dimensions, respectively.
When $s=1$, a UV-completion of the theory can be constructed \cite{Giombi:2019enr},
analogously to the higher-dimensional short-range $O(N)$ vector model \cite{Fei:2014yja}.

It has been argued that the IR
fixed point for $s>d/2$ is reached in the entire range $1<d<4$,
not merely in the perturbative regime around $d=2s-2\epsilon$.
Importantly, one is interested in the physically relevant integer dimension $d=3$ (where the spectrum of the theory is expected to be unitary), and general allowed values for
the exponent $s$. Some preliminary evidence in favor of this conclusion
is devised in the large-$N$ limit using the Hubbard-Stratonovich
transformation, analogously to the argument given in \cite{Gubser:2002vv} for the
short-range $O(N)$ vector models.
Specifically, at the Gaussian f.p. the composite singlet field $\phi^2$
is a primary operator of dimension $\Delta_{\phi^2}=2\Delta_\phi = d-s$.
As we will review momentarily, employing the Hubbard-Stratonovich transformation
one can demonstrate that the quartic interaction
brings the theory to a new regime where $\phi^2$
exhibits a scaling behavior with the (leading order in $1/N$) exponent $\Delta_{\phi^2} =\Delta_\s =s$,
where $\s$ is the Hubbard-Stratonovich field.

Using the Hubbard-Stratonovich transformation, the action (\ref{starting bi-local interaction action})
is equivalently re-written as
\begin{align}
 S= C(s)\int d^dx\,\int d^dy \f{\phi(x)\phi(y)}{|x-y|^{d+s}} + \int\,d^dx \left(-\f1{4g} \s^2 + \f1{\sqrt{N }} \s \phi^2\right)
 \label{eff-action-1}
\end{align}
The path integral over $\phi$ is now Gaussian, and integrating it out results in the effective action for $\s$
\begin{equation}
S = \frac{N}{2}\int d^dx\int d^dy\,\textrm{Tr}\,\log\left(\frac{C(s)}{|x-y|^{d+s}} + \frac{1}{\sqrt{N}}\,\sigma(x)
\delta^{(d)}(x-y)\right) - \frac{1}{4g}\int d^dx\,\sigma^2\,.
\label{eff-action-sigma}
\end{equation}
Using the inverse propagator relation
\begin{equation}
\label{inverse propagator}
\int d^dy\,\frac{1}{|y|^{2a}|x-y|^{2(d-a)}} = \pi^d A(a)A(d-a)\,\delta^{(d)}(x)\,,
\end{equation}
and expanding the logarithm we obtain
\begin{equation}
\label{Seff for s}
S = -C_\phi^2 \int d^dx\int d^dy\,\frac{\s(x)\s(y)}{|x-y|^{2(d-s)}} - \frac{1}{4g}\int d^dx\,\sigma^2 +{\cal O}\left(
\frac{1}{\sqrt{N}}\right)\,,
\end{equation}
where we also used the definition (\ref{Cphi def}).

For $s>d/2$, and in the long-distance IR limit, the first term in the r.h.s. of (\ref{Seff for s})
dominates over the second term.\footnote{Analogously, when $s<d/2$ the first term
dominates over the second term in the UV limit.} The second term then drops
out, and the $\langle\s\s\rangle$ propagator can be found from the first
term using (\ref{inverse propagator}).  As a result we obtain
\begin{equation}
\langle \s(x)\s(0)\rangle = \frac{C_\s}{|x|^{2\Delta_s}}\,,
\end{equation}
where the scaling dimension of the Hubbard-Stratonovich field $\s$
at the strongly-coupled regime in the IR is,
\begin{equation}
\label{Delta s}
\Delta_\s = s\,,
\end{equation}
and the propagator amplitude is found to be
\begin{align}
 C_\s= -\f{2^{2s-1}\G\left(d-s\right)\G\left(s\right)\G\left(\f{s}2\right)^2}{\G\left(\f{d}2-s\right)\G\left(s-\f{d}2\right)\G\left(\f{d-s}2\right)^2}\,.
\end{align}
As a consistency check, notice that for the scaling dimension (\ref{Delta s})
the second term in the r.h.s. of (\ref{Seff for s}) is irrelevant.

In the effective action (\ref{Seff for s})
we also omitted higher-order vertices for $\s$, which are suppressed in $1/N$.
We will encounter these vertices in our calculation of CFT data in this paper,
where they will be explicitly represented diagrammatically via polygon graphs
with internal $\phi$ propagator lines. These are constructed using the Feynman
rules:
\begin{center}
  \begin{picture}(300,67) (24,-17)
    \SetWidth{1.0}
    \SetColor{Black}
    \Line[](34,34)(98,34)
    \Vertex(34,34){2}
    \Vertex(98,34){2}
    \Text(106,23)[lb]{\scalebox{1.01}{$=\frac{C_\phi}{|x|^{2\Delta_\phi}}$}}
    \Text(106,-15)[lb]{\scalebox{1.01}{$=\frac{C_\s}{|x|^{2\Delta_\s}}$}}
    \Text(98,26)[lb]{\scalebox{0.81}{$x$}}
    \Text(24,26)[lb]{\scalebox{0.81}{$0$}}
    \Text(61,36)[lb]{\scalebox{0.81}{$2\Delta_\phi$}}
    \Line[](32,-6)(96,-6)
    \Text(21,-14)[lb]{\scalebox{0.81}{$0$}}
    \Text(98,-14)[lb]{\scalebox{0.81}{$x$}}
    \Text(60,-4)[lb]{\scalebox{0.81}{$2\Delta_\s$}}
    \Vertex(32,-6){2}
    \Vertex(96,-6){2}
    \Line[](185,34)(230,14)
    \Line[](230,14)(185,-6)
    \Line[](230,14)(272,14)
    \Vertex(230,14){4}
    \Text(289,5)[lb]{\scalebox{1}{$=-\frac{2}{\sqrt{N}}$}}
  \end{picture}
\end{center}
A propagator line with a generic exponent will be assumed to have a unit amplitude
\begin{center}
  \begin{picture}(300,7) (24,32)
    \SetWidth{1.0}
    \SetColor{Black}
    \Line[](34,34)(98,34)
    \Vertex(34,34){2}
    \Vertex(98,34){2}
    \Text(61,36)[lb]{\scalebox{0.81}{$2a$}}
    \Text(98,26)[lb]{\scalebox{0.81}{$x$}}
    \Text(24,26)[lb]{\scalebox{0.81}{$0$}}
    \Text(106,25)[lb]{\scalebox{1.01}{$=\frac{1}{|x|^{2a}}$}}
    \end{picture}
\end{center}
In appendix~\ref{app_a} we collect some well-known 
identifies for conformal graphs in position space, that have been
used for calculations in this paper.

\section{Two-point functions}
\label{sec:two-pt}

In this section we will calculate the $\langle \phi\phi\rangle$ and $\langle \sigma\sigma\rangle$
two-point functions of the fundamental field $\phi$ and the Hubbard-Stratonovich
scalar field $\sigma$. Conformal invariance ensures that these two-point functions have the form
\begin{align}
\label{<phi phi> general}
\langle \phi(x)\phi(0)\rangle &= \frac{C_\phi(1+A_\phi)\mu^{-2\gamma_\phi}}{|x|^{2(\Delta_\phi+\gamma_\phi)}}\,,\\
\label{<sigma sigma> general}
\langle \sigma(x)\sigma(0)\rangle &= \frac{C_\sigma(1+A_\sigma)\mu^{-2\gamma_\sigma}}{|x|^{2(\Delta_\sigma+\gamma_\sigma)}}\,,
\end{align}
where $\mu$ is an arbitrary mass scale, and $\Delta_{\phi, \s}$ are leading order dimensions at the
long-range critical point, defined in section~\ref{sec:setup}.
In this section we will calculate the two-point functions (\ref{<phi phi> general}),
(\ref{<sigma sigma> general}) at the next-to-leading order in $1/N$ expansion.
In particular, we will reproduce the known expression for the anomalous dimensions
$\gamma_{\sigma}$, and obtain new results for the relative corrections $A_{\phi,\sigma}$
to the amplitudes of these two-point functions.

We will also demonstrate that the anomalous dimension $\gamma_\phi$ vanishes
at the next-to-leading order in $1/N$. As discussed in section~\ref{sec:setup}, in fact
we expect the engineering dimensions $\Delta_\phi$ of the field $\phi$ to be exact to all
orders in $1/N$, being fixed by the non-local kinetic term. However, our calculation
of the two-point function  $\langle \phi\phi\rangle$ is still useful, because we extract from
it the relative amplitude correction $A_\phi$. While the propagator
amplitude corrections are not observables, together with its counterpart $A_\sigma$, it will
play an important part
 in our calculation in section~\ref{sec:OPE coefficients} of the observable CFT data, such as the amplitudes
of the $\langle \phi\phi\sigma\rangle$ and
$\langle \s\s\sigma\rangle$ three-point functions and the related $\phi\phi$ and $\s\s$ OPE coefficients.

\subsection{$\lan\phi\phi\ran$}
\label{sec: phi phi correction}

It is easy to see that the only diagram contributing to the $\langle\phi\phi\rangle$ two-point 
function at the next-to-leading order in $1/N$ is given by
\begin{center}
  \begin{picture}(670,65) (100,-35)
    \Vertex(180,-4){2}
    \Line[](180,-4)(230,-4)
    \Text(175,0)[lb]{\scalebox{0.8}{$0$}}
    \Vertex(230,-4){4}
    \Arc[](255,-4)(25,-180,-360)
    \Arc[](255,-4)(25,0,180)
    \Vertex(280,-4){4}
    \Line[](280,-4)(330,-4)
    \Text(325,0)[lb]{\scalebox{0.8}{$x$}}
    \Vertex(330,-4){2}
    \Text(195,0)[lb]{\scalebox{0.8}{$2\Delta_\phi$}}
    \Text(250,-42)[lb]{\scalebox{0.8}{$2\Delta_\phi$}}
    \Text(240,25)[lb]{\scalebox{0.8}{$2\Delta_\s +\delta$}}
    \Text(300,0)[lb]{\scalebox{0.8}{$2\Delta_\phi$}}
    \Text(350,-15)[lb]{$=\frac{C_\phi\,\mu^{-\delta}}{|x|^{2\Delta_\phi +\delta}}\,\left(\frac{2\gamma_\phi}{\delta} + A_\phi
    +{\cal O}(\delta)\right)$}
  \end{picture}
\end{center}
Here we have introduced an auxiliary regulator $\delta$ to the internal $\s$
line. Evaluating the diagram gives\footnote{In this paper
we utilize the \textit{supset} sign $\supset$ to denote some of the terms which contribute
to the expression on the l.h.s. of this sign.}
\begin{align}
\langle \phi(x)\phi(0)\rangle\supset\f4N \f{C_\phi^3 C_\s}{|x|^{2\df}} U\left(\f{d{-}s}2,\f{d{+}s{+}\d}2,{-}\f{\d}2\right)U\left(\f{d{+}\d}2,\f{d{-}s}2,\f{s{-}\d}2\right)\f1{(\mu|x|)^\d}\,.
\end{align}
Expanding this expression around $\delta = 0$ shows that it is in fact finite and $\gamma_\phi = 0$,\footnote{A simple explanation of the vanishing anomalous dimension, $\gamma_\phi = 0$, is that only local divergences can appear from the loops, which in turn are to be cancelled by local counter-terms. However, since the fields $\phi_i$ have a bi-local action, no local counter-terms can be obtained by the wave function renormalization of the $\phi_i$.
Correspondingly, no divergences can appear in the loops of this renormalizable theory.
In other words, the loops cannot induce wave-function renormalization of $\phi_i$.}
while
\begin{align}
\label{Aphi}
A_\phi = \frac{1}{N}\frac{2^{d-s} (2 s-d)  \Gamma (s) \sin \left(\frac{1}{2} \pi  (d-2 s)\right) \Gamma \left(\frac{1}{2} (d-s+
1)\right)}{\sqrt{\pi } s\Gamma \left(\frac{d+s}{2}\right)\sin \left(\frac{\pi  s}{2}\right)} + {\cal O}\left(\frac{1}{N^2}\right)\,.
\end{align}

\subsection{$\lan\s\s\ran$}
\label{sec: <ss>}

There are three diagrams that contribute to the $\langle\s\s\rangle$ correlation
function at the next-to-leading order in $1/N$, which we will label as $C_{\s\s}^{(a)}$, $a=1,2,3$.
Just as in the case of  $\langle\phi\phi\rangle$ two-point function, these diagrams 
are analogous to their counterparts in the short-range critical $O(N)$ vector model,
and therefore calculation sequences in these two models parallel each other.
We refer the reader to \cite{Goykhman:2019kcj} for a recent detailed description of such
a calculation in the $O(N)$ vector model, while here we provide only a brief outline and a summary of results.

The total values of the anomalous dimension and the propagator amplitude correction 
at the next-to-leading order in $1/N$
are obtained by summing the individual contributions from each of the three diagrams,
\begin{equation}
\label{gamma s and As}
\gamma_\s = \sum_{a=1}^3 \gamma_\s^{(a)}\,,\qquad
A_\s = \sum_{a=1}^3 A_\s^{(a)}\,.
\end{equation}

The first diagram contributing to $\langle\s\s\rangle$ is given by
\begin{center}
  \begin{picture}(194,90) (85,-37)
    \SetWidth{1.0}
    \SetColor{Black}
    \Text(-20,7)[lb]{$C_{\s\s}^{(1)}:$}
    \Text(245,2)[lb]{$=\frac{C_\s\,\mu^{-\delta}}{|x|^{2\Delta_\s+\delta}}\,\left(\frac{2\gamma_\s^{(1)}}{\delta} + A_\s^{(1)}
    +{\cal O}(\delta)\right)$}
    \Line[](32,10)(94,10)
    \Text(55,15)[lb]{\scalebox{0.6}{$2\Delta_\s$}}
    \Arc[](128,10)(35,153,513)
    \Line[](162,10)(224,10)
    \Text(190,15)[lb]{\scalebox{0.6}{$2\Delta_\s$}}
    \Line(128,44)(128,-24)
    \Text(133,15)[lb]{\scalebox{0.6}{$2\Delta_\s+\delta$}}
    \Text(73,35)[lb]{\scalebox{0.6}{$2\Delta_\phi-\eta$}}
    \Text(73,-20)[lb]{\scalebox{0.6}{$2\Delta_\phi+\eta$}}
    \Text(157,-20)[lb]{\scalebox{0.6}{$2\Delta_\phi+\eta$}}
    \Text(157,35)[lb]{\scalebox{0.6}{$2\Delta_\phi-\eta$}}
    \Vertex(32,10){2}
    \Vertex(94,10){4}
    \Vertex(162,10){4}
    \Vertex(224,10){2}
    \Vertex(128,-24){4}
    \Vertex(128,44){4}
    \Text(20,5)[lb]{\scalebox{0.8}{$0$}}
    \Text(230,5)[lb]{\scalebox{0.8}{$x$}}
  \end{picture}
\end{center}
Here an auxiliary regulator $\eta={\cal O}(\delta)$ does not affect the value of $C_{\s\s}^{(1)}$ in the $\delta\rightarrow 0$
limit \cite{Vasiliev:1981yc,Vasiliev:1981dg,Gubser:2017vgc}. Choosing $\eta=\delta/2$ makes the diagram
integrable via the uniqueness relation, rendering
\begin{align}
\langle \sigma(x)\sigma(0)\rangle
 &\supset\f8{N}\f{C_\phi^4C_\s^3}{|x_{12}|^{2\ds}}U\left(\df-\f{\d}4,\df-\f{\d}4,\ds+\f{\d}2\right)U\left(\f{d+\d}2,\f{d+\d}2,-\d\right)\nn\\
 &\times U\left(\ds,2\df+\f{\d}2,-\f{\d}2\right)U\left(\ds,\f{d+\d}2,\f{d-\d}2-\ds\right)\,
 \f1{(\mu|x|)^\d}\,.
\end{align}
Expanding around $\delta = 0$ one obtains the corresponding contributions
of the diagram $C_{\s\s}^{(1)}$ to the $\gamma_\s$ and $A_\s$,\footnote{In this paper we denote $n$th
derivative of the digamma function as $\psi^{(n)}(x)$.}
\begin{align}
\label{gammas1}
\gamma_\s^{(1)} &= - \frac{1}{N}\,\frac{4 \Gamma \left(\frac{s}{2}\right)^2 \Gamma (d-s)}{  \Gamma \left(\frac{d}{2}\right) \Gamma \left(\frac{d-s}{2}\right)^2 \Gamma \left(s-\frac{d}{2}\right)} \,,\\
\label{As1}
A_\s^{(1)} &= \gamma_\s^{(1)}
\left(\psi ^{(0)}\left(\frac{d{-}s}{2}\right)-\psi ^{(0)}(d{-}s)
\psi ^{(0)}\left(s{-}\frac{d}{2}\right)+\psi ^{(0)}\left(\frac{s}{2}\right)\right)\,.
\end{align}

Next, we consider contribution of the following diagram to $\langle \s\s\rangle$,
\begin{center}
  \begin{picture}(300,130) (31,-65)
    \SetWidth{1.0}
    \SetColor{Black}
    \Text(-25,11)[lb]{$C_{\s\s}^{(2)}:$}
    \Text(55,22)[lb]{\scalebox{0.8}{$2\Delta_\s$}}
    \Text(320,22)[lb]{\scalebox{0.8}{$2\Delta_\s$}}
    \Text(70,-22)[lb]{\scalebox{0.8}{$2\Delta_\phi+\eta$}}
    \Text(285,-22)[lb]{\scalebox{0.8}{$2\Delta_\phi-\eta$}}
    \Text(280,45)[lb]{\scalebox{0.8}{$2\Delta_\phi+\eta$}}
    \Text(70,45)[lb]{\scalebox{0.8}{$2\Delta_\phi-\eta$}}
    \Text(180,55)[lb]{\scalebox{0.8}{$2\Delta_\s+\delta/2$}}
    \Text(180,-30)[lb]{\scalebox{0.8}{$2\Delta_\s+\delta/2$}}
    \Text(140,11)[lb]{\scalebox{0.8}{$2\Delta_\phi$}}
    \Text(230,11)[lb]{\scalebox{0.8}{$2\Delta_\phi$}}
    \Text(20,11)[lb]{$0$}
    \Text(355,5)[lb]{$x$}
    \Line[](32,16)(94,16)
    \Arc[](128,16)(35,153,513)
    \Line[](144,48)(240,48)
    \Line[](144,-16)(240,-16)
    \Arc[](256,16)(35,153,513)
    \Line[](292,16)(354,16)
    \Vertex(32,16){2}
    \Vertex(94,16){4}
    \Vertex(292,16){4}
    \Vertex(354,16){2}
    \Vertex(144,48){4}
    \Vertex(240,48){4}
    \Vertex(144,-16){4}
    \Vertex(240,-16){4}
    \Text(0,-65)[lb]{$=\frac{C_\s\,\mu^{-\delta}}{|x|^{2\Delta_\s+\delta}}\,\left(\frac{2\gamma_\s^{(2)}}{\delta} + A_\s^{(2)}
    +{\cal O}(\delta)\right)$}
  \end{picture}
\end{center}
Here we again introduced an auxiliary regulator $\eta = {\cal O}(\delta/2)$ without affecting the 
value of this diagram in the $\delta\rightarrow 0$ limit. We choose $\eta = \delta/2$,
which renders some of the vertices unique and therefore integrable. Integrating over those
vertices will result in a graph of the same topology to $C_{\s\s}^{(1)}$,
which will require introducing yet again an auxiliary regulator $\eta' = {\cal O}(\delta)$,
in such a manner that the final answer in the $\delta\rightarrow 0$ limit will remain unaffected,
while the uniqueness of the diagram becomes restored \cite{Vasiliev:1981yc,Vasiliev:1981dg,Gubser:2017vgc}. At the end, we obtain
\begin{align}
\langle \sigma(x)\sigma(0)\rangle &\supset \f{32}{N}\f{C_\phi^6C_\s^4}{|x_{12}|^{2\ds}}U\left(\df-\f{\d}4,\df,\ds+\f{\d}4\right)^2U\left(d-\f{3\ds}2-\f{\d}4,\f{\ds}2-\f{\d}4,\ds+\f{\d}2\right)\\
 &\times U\left(\f{d{+}\d}2,\f{d{+}\d}2,{-}\d\right)U\left(2\df{+}\f{\d}2,\ds,{-}\f{\d}2\right)
 U\left(\ds,\f{d{+}\d}2,\f{d{-}\d}2{-}\ds\right)\f1{(\mu|x|)^\d}\,.\nn
\end{align}
The contributions to anomalous dimension and the propagator amplitude are then given by
\begin{align}
\label{gammas2}
\gamma_\s^{(2)} &= \frac{1}{N}\,\frac{16 \Gamma \left(\frac{s}{2}\right)^3 \Gamma \left(\frac{d}{2}-s\right) \Gamma (d-s)^2 \Gamma \left(\frac{3 s}{2}-\frac{d}{2}\right)}{  \Gamma \left(\frac{d}{2}\right) \Gamma (s) \Gamma \left(d-\frac{3 s}{2}\right) \Gamma \left(\frac{d-s}{2}\right)^3 \Gamma \left(s-\frac{d}{2}\right)^2}\,,\\
\label{As2}
A_\s^{(2)} &{=} \frac{\gamma_\s^{(2)}}{4}\,\left(3 \psi ^{(0)}\left(\frac{d{-}s}{2}\right)
{+}\psi ^{(0)}\left(\frac{3 s}{2}
{-}\frac{d}{2}\right){+}\psi ^{(0)}\left(d{-}\frac{3 s}{2}\right){-}2 \psi ^{(0)}\left(\frac{d}{2}{-}s\right)\right.\\
&{-}\left.2 \psi ^{(0)}(d{-}s){-}2 \left(\psi ^{(0)}\left(s{-}\frac{d}{2}\right){+}\psi ^{(0)}(s)\right){+}3 \psi ^{(0)}\left(\frac{s}{2}\right)\right)\,.\notag
\end{align}

Finally, the third diagram contributing to $\langle\s\s\rangle$ 
at the next-to-leading order in $1/N$ is obtained by incorporating $1/N$ correction to
the $\langle\phi\phi\rangle$ sub-diagram of the leading-order $\phi$ bubble:
\begin{center}
  \begin{picture}(194,90) (81,-27)
    \SetWidth{1.0}
    \SetColor{Black}
    \Text(-30,7)[lb]{$C_{\s\s}^{(3)}:$}
    \Text(245,0)[lb]{$=\frac{C_\s\,\mu^{-\delta}}{|x|^{2\Delta_\s+\delta}}\,\left(\frac{2\gamma_\s^{(3)}}{\delta} + A_\s^{(3)}
    +{\cal O}(\delta)\right)$}
    \Text(16,5)[lb]{\scalebox{0.8}{$0$}}
    \Line[](32,10)(94,10)
    \Text(52,15)[lb]{\scalebox{0.8}{$2\Delta_\s$}}
    \Text(73,15)[lb]{\scalebox{0.8}{$2\Delta_\phi$}}
    \Text(168,15)[lb]{\scalebox{0.8}{$2\Delta_\phi$}}
    \Text(233,5)[lb]{$x$}
    \Arc[](128,10)(35,153,513)
    \Line[](162,10)(224,10)
    \Text(195,15)[lb]{\scalebox{0.8}{$2\Delta_\s$}}
    \Line[](97,25)(159,25)
    \Text(115,15)[lb]{\scalebox{0.8}{$2\Delta_\s+\delta$}}
    \Text(125,48)[lb]{\scalebox{0.8}{$2\Delta_\phi$}}
    \Text(125,-36)[lb]{\scalebox{0.8}{$2\Delta_\phi$}}
    \Vertex(32,10){2}
    \Vertex(94,10){4}
    \Vertex(162,10){4}
    \Vertex(224,10){2}
    \Vertex(97,25){4}
    \Vertex(159,25){4}
  \end{picture}
\end{center}
Such a correction to the $\phi$ propagator sub-diagram has been discussed
in section~\ref{sec: phi phi correction}, where it was established to be finite.
Specifically, we obtain
\begin{align}
\langle \sigma(x)\sigma(0)\rangle&\supset\f{16}N \f{C_\phi^4 C_\s^3}{|x|^{2\ds}}U\left(\df,\df+\ds+\f{\d}2,-\f{\d}2\right)U\left(\df,\f{d+\d}2,\f{d-\d}2-\df\right)\nn\\
&U\left(2\df+\f{\d}2,\ds,-\f{\d}2\right)U\left(\ds,\f{d+\d}2,\f{d-\d}2-\ds\right)\f1{(\mu|x|)^\d}\,,
\end{align}
expanding which around $\delta = 0$ gives
\begin{align}
\label{gammas3}
\gamma_\s^{(3)} &= 0\,,\\
\label{As3}
A_\s^{(3)} &{=} \frac{1}{N}\,\frac{2^{d-s+1} (d-2 s) \csc \left(\frac{\pi  s}{2}\right) \Gamma (s) \sin \left(\frac{1}{2} \pi  (d-2 s)\right) \Gamma \left(\frac{1}{2} (d-s+1)\right)}{\sqrt{\pi } s \Gamma \left(\frac{d+s}{2}\right)}\,.
\end{align}

Combining (\ref{gamma s and As}), (\ref{gammas1}), (\ref{As1}), (\ref{gammas2}), (\ref{As2}), (\ref{gammas3}), (\ref{As3})
we obtain the final answer for $\gamma_\s$, $A_\s$. Our result for $\gamma_\s$,
\begin{align}
\notag
\gamma_\s&{=}-\frac{1}{N}\frac{4 \Gamma \left(\frac{s}{2}\right)^2 \Gamma (d{-}s)}
{\Gamma \left(\frac{d}{2}\right) \Gamma (s) \Gamma
\left(d{-}\frac{3 s}{2}\right) \Gamma \left(\frac{d{-}s}{2}\right)^3 \Gamma \left(s{-}\frac{d}{2}\right)^2}
\left(\Gamma (s) \Gamma \left(d{-}\frac{3 s}{2}\right) \Gamma \left(\frac{d{-}s}{2}\right) \Gamma \left(s{-}\frac{d}{2}\right)\right.\\
&-\left. 2 \Gamma \left(\frac{s}{2}\right) \Gamma \left(\frac{d}{2}-s\right)
\Gamma (d-s) \Gamma \left(\frac{3 s}{2}-\frac{d}{2}\right)\right)\,,\label{gamma_sigma result}
\end{align}
agrees with \cite{Gubser:2017vgc,Giombi:2019enr}, while the expressions we obtained for the relative corrections $A_{\phi,\s}$
to the propagators of the $\phi$  and $\s$ are, to the best of our knowledge, new. While
these are not observables, in combination with the $1/N$ corrections to the effective
cubic interaction vertices (conformal triangles) such as $\phi\phi\s$ and $\s\s\s$,
they give amplitudes of the three-point functions (\textit{e.g.}, $\langle\phi\phi\s\rangle$
and $\langle\s\s\s\rangle$), which determine the OPE coefficients, and are a part of CFT data.

\section{Conformal invariance and composite operators}
\label{sec:conf-inv}

In this section, we will discuss conformal invariance of the long-range fixed point at the level of two-point correlation functions of composite operators such as $\sigma^n$, $n>1$, and $\sigma\phi$. Recall that in the short-range critical $O(N)$ vector model composite operators of the kind $\hat \s^n$, $n>1$\footnote{We use $\hat \s$ to denote the Hubbard-Stratonovich field in the short-range critical 
$O(N)$ vector model, while its counterpart in the long-range model is denoted as $\s$.} are usually scaling operators \cite{Derkachov:1997gc,Derkachov:1998js}.\footnote{See \cite{Ma:1974qh} for an extensive earlier work on the formalism of scaling operators.} These operators, therefore, have a fixed scaling dimension, and further acquire an anomalous dimension at the fixed point. However, such operators are usually not conformal primaries in the theory. In fact, since $\hat\s$ has a fixed scaling dimension, $\Delta_{\hat\s}=2$ for any $2<d<4$, it mixes with other descendent operators of the same dimension, created by replacing some of the $\hat\s$ fields with derivatives. The resulting operators furnish true conformal primaries in the theory \cite{Derkachov:1997gc,Derkachov:1998js}.

The spectrum of long-range CFTs is slightly different. The Hubbard-Stratonovich field $\s$ in such a theory has a leading scaling dimension $\Delta_\s=s$, which is generally a non-integer number. It implies immediately that one cannot create local descendent operators with leading-order scaling dimensions equal to that of $\s^n$, by simply replacing some of the $\s$ fields with derivatives. Such composite operators, therefore, do not mix with any other operators, and are expected to be conformal primaries by themselves.\footnote{The case of $s=1$ needs a separate discussion.} Below, we will test these statements at the level of two-point correlators of some composite operators. If the composite operators are indeed conformal primaries
in the long-range CFT, we will find that the two point function of any composite operator with itself exhibits a power-law scaling behavior, while its correlators with other operators of different scaling dimensions are exactly zero. 

\subsection{$\lan \s^2 \s^2\ran$}
\label{sec:s2s2}

We begin by calculating the two-point correlation function of the composite operator $\s^2$.
Starting from the assumption that $\s^2$ is a scaling operator, as we will explicitly establish below
at the next-to-leading order in $1/N$ expansion, its two-point function is expected to have 
the form
\begin{align}
 \lan \s^2(x)\s^2(0)\ran = \f{C_{\s^2}(1+A_{\s^2})}{|x|^{2(\Delta_{\s^2}+\g_{\s^2})}}\,,
\end{align}
where we separated the amplitude into the leading part $C_{\s^2}$ and its relative
$1/N$ corrections $A_{\s^2}$. At the same time, $\Delta_{\s^2}$ stands for the leading
order scaling dimension, while $\g_{\s^2}$ is the anomalous dimension contribution.
In the large-$N$ limit the $\lan\s^2\s^2\ran$ propagator is determined by the diagram
\begin{center}
  \begin{picture}(368,60) (37,10)
    \SetWidth{1.0}
    \SetColor{Black}
    \CBox(45,35)(50,40){Black}{Black}
    \Arc[clock](287.5,-32.976)(91.978,129.496,50.504)
    \Arc[](287.5,117.5)(98.501,-127.161,-52.839)
    \CBox(169,35)(174,40){Black}{Black}
    \CBox(224,35)(229,40){Black}{Black}
    \CBox(345,35)(350,40){Black}{Black}
    \Line[](50,37)(169,37)
    \Text(22,24)[lb]{\scalebox{0.81}{$\sigma(0)^2$}}
    \Text(166,24)[lb]{\scalebox{0.81}{$\sigma(x)^2$}}
    \Text(105,43)[lb]{\scalebox{0.81}{$2\Delta_{\s^2}$}}
    \Text(200,36)[lb]{\scalebox{0.81}{$=$}}
    \Text(285,64)[lb]{\scalebox{0.81}{$2\Delta_\s$}}
    \Text(285,7)[lb]{\scalebox{0.81}{$2\Delta_\s$}}
    \Text(370,26)[lb]{\scalebox{1.21}{$=\frac{C_{\s^2}}{|x|^{2\Delta_{\s^2}}}$}}
    \Text(208,21)[lb]{\scalebox{0.81}{$\sigma(0)^2$}}
    \Text(350,21)[lb]{\scalebox{0.81}{$\sigma(x)^2$}}
  \end{picture}
\end{center}
from which one easily finds
\begin{align}
\label{leading order s2 prop amplitude and dim}
 C_{\s^2}= 2C_\s^2\,,\qquad \Delta_{\s^2}=2\Delta_\s=2s\,.
\end{align}

There are four diagrams that contribute to the $\lan\s^2\s^2\ran$ propagator at the next-to-leading order in $1/N$.
The first contribution is simply due to the $1/N$ corrections to the propagators of $\s$ in the leading order diagram.
We calculated these corrections in section~\ref{sec: <ss>}, and will denote their total contribution with a gray blob: 
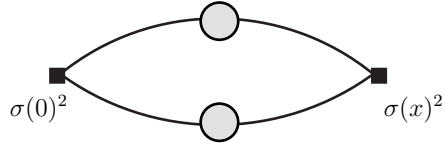
\begin{center}
  \begin{picture}(368,65) (-70,10)
    \SetWidth{1.0}
    \SetColor{Black}
    \CBox(45,35)(50,40){Black}{Black}
    \Arc[clock](107,-32.976)(91.978,129.496,50.504)
    \Arc[](107,117.5)(98.501,-127.161,-52.839)
    \GOval(108,58)(7,7)(0){0.882}
    \GOval(108,20)(7,7)(0){0.882}
    \CBox(165,35)(170,40){Black}{Black}
    \Text(30,20)[lb]{\scalebox{0.81}{$\sigma(0)^2$}}
    \Text(170,20)[lb]{\scalebox{0.81}{$\sigma(x)^2$}}
  \end{picture}
\end{center}
This diagram includes the leading-order $\lan\s^2\s^2\ran$ propagator, as well
as some contributions to $\g_{\s^2}$ and $A_{\s^2}$,
\begin{align}
\label{1st contribution to <s^2s^2>}
 \lan \s^2(x)\s^2(0)\ran\supset \f{C_{\s^2}(1+A_\s)^2}{|x|^{4\Delta_{\s}}}\left(1-4\g_\s\log(|x|\mu)\right)\,.
\end{align}
Here, and in what follows, we are going to ignore the finite terms such as $2A_\s$ in (\ref{1st contribution to <s^2s^2>}),
and focus only on the singular terms, that contribute to the anomalous
dimension $\gamma_{\s^2}$. This will greatly simplify the calculation,
since we can avoid introducing the regulator $\delta$, and therefore easily
take all of the unique integrals. The last logarithmically divergent
integral is simply replaced with $(2\pi^\frac{d}{2}/\Gamma(d/2))\,\log(\mu)$,
where $\mu$ is IR mass scale \cite{Goykhman:2020ffn,Chai:2020hnu}.

The second contribution to $\lan\s^2\s^2\ran$
at the next-to-leading order in $1/N$ is due to insertions
of two cubic $\s$ vertices into the leading order diagram:\footnote{To lighten up the notation, we skip
labeling exponents on diagrams in the rest of this subsection.}
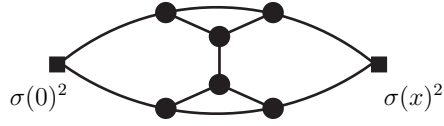
\begin{center}
  \begin{picture}(368,65) (-70,10)
    \SetWidth{1.0}
    \SetColor{Black}
    \CBox(45,35)(50,40){Black}{Black}
    \Arc[clock](107,-32.976)(91.978,129.496,50.504)
    \Arc[](107,117.5)(98.501,-127.161,-52.839)
    \Line[](108,30)(108,48)
    \Line[](108,30)(88,21)
    \Line[](108,30)(128,21)
    \Line[](108,48)(88,57)
    \Line[](108,48)(128,57)
    \Vertex(108,30){4}
    \Vertex(108,48){4}
    \Vertex(88,21){4}
    \Vertex(88,57){4}
    \Vertex(128,21){4}
    \Vertex(128,57){4}
    \CBox(165,35)(170,40){Black}{Black}
    \Text(30,20)[lb]{\scalebox{0.81}{$\sigma(0)^2$}}
    \Text(170,20)[lb]{\scalebox{0.81}{$\sigma(x)^2$}}
  \end{picture}
\end{center}
Evaluating this diagram gives
\begin{align}
\label{2nd contribution to <s^2s^2>}
 \lan \s^2(x)\s^2(0)\ran &\supset \f{128}N\f{C_\phi^6 C_\s^5}{|x|^{2\Delta_{\s^2}}} U\left(s,\f{d-s}2,\f{d-s}2\right)^2 U\left(\f{s}2,s,d-\f{3s}2\right)\\
 &\times U\left(\f{d-s}2,d-\f{3s}2,2s-\f{d}2\right)U\left(s,s,d-2s\right)\f{4\pi^{d/2}}{\G\left(d/2\right)}\log(|x|\mu)\,.\nn
\end{align}
The third and fourth diagrams are
obtained due to insertion of a single quartic vertex into the propagator.
The corresponding planar diagram is given by
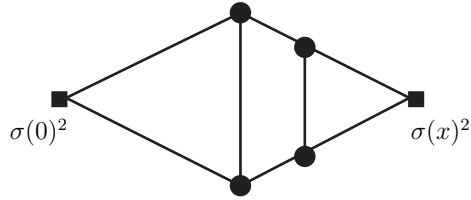
\begin{center}
  \begin{picture}(305,67) (-70,-16)
    \SetWidth{1.0}
    \SetColor{Black}
    \Line[](21,18)(86,50)
    \Line[](149,18)(86,50)
    \Vertex(86,50){4}
    \Line[](21,18)(86,-15)
    \Line[](86,-15)(149,18)
    \Vertex(86,-15){4}
    \Line[](86,50)(86,-15)
    \Line[](110,-3)(110,37)
    \Vertex(110,37){4}
    \Vertex(110,-4){4}
    \CBox(16,15)(21,20){Black}{Black}
    \CBox(149,15)(154,20){Black}{Black}
    \Text(0,0)[lb]{\scalebox{0.81}{$\sigma(0)^2$}}
    \Text(150,0)[lb]{\scalebox{0.81}{$\sigma(x)^2$}}
   \end{picture}
\end{center}
and contributes
\begin{align}
\label{3rd contribution to <s^2s^2>}
\lan \s^2(x)\s^2(0)\ran\supset \f{32}{N}\f{C_\phi^4C_\s^4}{|x|^{2\Delta_{\s^2}}}U\left(s,\f{d{-}s}2,\f{d{-}s}2\right)^2 U\left(\f{s}2,\f{3s}2,d{-}2s\right)\f{4\pi^{d/2}}{\G\left(d/2\right)}\log(|x|\mu)\,,
\end{align}
while the non-planar diagram
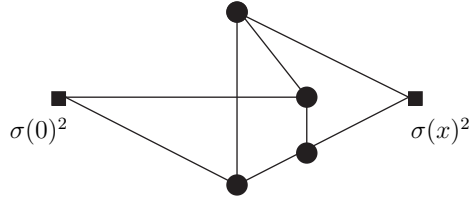
\begin{center}
  \begin{picture}(305,67) (110,-16)
    \Line[](200,18)(264,-15)
    \Vertex(264,-15){4}
    \Line[](264,50)(328,18)
    \Vertex(264,50){4}
    \Line[](264,-15)(328,18)
    \Line[](264,50)(264,-15)
    \Vertex(264,50){4}
    \Line[](200,18)(290,18)
    \Vertex(290,18){4}
    \Line[](290,-3)(290,18)
    \Vertex(290,-3){4}
    \Line[](290,18)(265,50)
    \CBox(195,15)(200,20){Black}{Black}
    \CBox(328,15)(333,20){Black}{Black}
    \Text(180,0)[lb]{\scalebox{0.81}{$\sigma(0)^2$}}
    \Text(330,0)[lb]{\scalebox{0.81}{$\sigma(x)^2$}}
  \end{picture}
\end{center}
contributes
\begin{align}
\label{4th contribution to <s^2s^2>}
 \lan \s^2(x)\s^2(0)\ran\supset\f{16}{N}\f{C_\phi^4C_\s^4}{|x|^{2\Delta_{\s^2}}}U\left(s,\f{d{-}s}2,\f{d{-}s}2\right)^2
 U\left(s,s,d{-}2s\right)
 \f{4\pi^{d/2}}{\G\left(d/2\right)}\log(|x|\mu)\,.
\end{align}

Combining (\ref{1st contribution to <s^2s^2>}), (\ref{2nd contribution to <s^2s^2>}), (\ref{3rd contribution to <s^2s^2>}) and
(\ref{4th contribution to <s^2s^2>}) we obtain the anomalous dimension of $\s^2$
at the next-to-leading order in $1/N$,
\begin{align}
 &\g_{\s^2}=-\f1{N}\frac{4 \Gamma \left(\frac{s}{2}\right)^2 \Gamma (d-s)}{
   \Gamma \left(\frac{d}{2}\right) \Gamma (s)^2 \Gamma \left(\frac{3
   s}{2}\right) \Gamma (d-2 s) \Gamma \left(d-\frac{3 s}{2}\right)^2 \Gamma
   \left(\frac{d-s}{2}\right)^4 \Gamma \left(s-\frac{d}{2}\right)^3}\nn\\
   & \times\Bigg(\Gamma
   \left(\frac{s}{2}\right) \Gamma (d-s)\Gamma \left(s-\frac{d}{2}\right) \Gamma
   \left(2 s-\frac{d}{2}\right) \Gamma \left(d-\frac{3 s}{2}\right)^2\nn\\
   &\times\left(\Gamma \left(\frac{s}{2}\right)
   \Gamma \left(\frac{3 s}{2}\right) \Gamma \left(\frac{d}{2}-s\right)^2+2
   \Gamma (s)^2 \Gamma \left(\frac{1}{2} (d-3 s)\right) \Gamma
   \left(\frac{d-s}{2}\right)\right) \nn\\
   &+2 \Gamma
   \left(\frac{3 s}{2}\right) \Gamma (d-2 s) \Bigg(\Gamma (s)^2 \Gamma
   \left(d-\frac{3 s}{2}\right)^2 \Gamma \left(\frac{d-s}{2}\right)^2 \Gamma
   \left(s-\frac{d}{2}\right)^2\nn\\
   &-2 \Gamma \left(\frac{s}{2}\right) \Gamma (s)
   \Gamma \left(d-\frac{3 s}{2}\right) \Gamma \left(\frac{d}{2}-s\right) \Gamma
   \left(\frac{d-s}{2}\right) \Gamma (d-s) \Gamma \left(\frac{3
   s}{2}-\frac{d}{2}\right) \Gamma \left(s-\frac{d}{2}\right)\nn\\
   &-2 \Gamma
   \left(\frac{s}{2}\right)^2 \Gamma \left(\frac{d}{2}-s\right)^2 \Gamma
   (d-s)^2 \Gamma \left(\frac{3 s}{2}-\frac{d}{2}\right)^2\Bigg)\Bigg)
\end{align}

A quick consistency check of this expression is given by 
$\g_{\s^2}|_{s=d/2} = 0$, since at $s=d/2$ the long-range CFT becomes free.
Another check is given by continuity at the long-range--short-range crossover point $s=s_\star$,
which demands $\gamma_{\s^2}|_{s=s_\star} = \gamma_{\hat\s^2}$,
where $\gamma_{\hat\s^2}$ is anomalous
dimension of the composite operator $\hat\s^2$ in short-range $O(N)$ vector model CFT.
This crossover will be discussed in more detail in section~\ref{sec: crossover to short range}.


\subsection{$\lan \sigma^n \sigma^n \ran$}
The $\lan \sigma^n \sigma^n \ran$ propagator
of the composite field $\sigma^n$ (where  $n \in \mathbb{N}$)
at the next-to-leading order in $1/N$ is determined by the diagrams
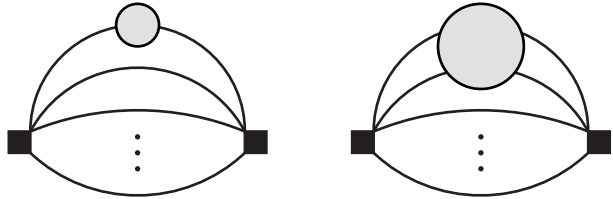
\begin{center}
  \begin{picture}(500,79) (31,-15)
    \SetWidth{1.0}
    \SetColor{Black}
    \CBox(160,7)(168,15){Black}{Black}
    \Arc[](208,49)(58,-133.603,-46.397)
    \CBox(248,7)(256,15){Black}{Black}
    \Arc[clock](208,-81)(104,112.62,67.38)
    \Arc[clock](208,-6.333)(45.333,151.928,28.072)
    \Arc[clock](208,15)(40,-180,-360)
    \Arc[](336,49)(58,-133.603,-46.397)
    \CBox(288,7)(296,15){Black}{Black}
    \CBox(376,7)(384,15){Black}{Black}
    \Arc[clock](336,-81)(104,112.62,67.38)
    \Arc[clock](336,-6.333)(45.333,151.928,28.072)
    \Arc[clock](336,15)(40,-180,-360)
    \GOval(336,47)(16,16)(0){0.882}
    \GOval(208,55)(8,8)(0){0.882}
    \Vertex(208,1){1}
    \Vertex(208,7){1}
    \Vertex(208,13){1}
    \Vertex(336,1){1}
    \Vertex(336,7){1}
    \Vertex(336,13){1}
  \end{picture}
\end{center}
Here we used gray blobs to denote dressed $\langle\sigma\sigma\rangle$ 
and $\langle\sigma^2\sigma^2\rangle$ propagators.
The anomalous dimension $\gamma_{\sigma^n}$ is therefore determined only by $\gamma_\sigma$ and $\gamma_{\sigma^2}$. Further corrections contribute only at higher orders in $1/N$. The value of $\gamma_{\sigma^n}$ is then found to be,
\begin{equation}
\gamma_{\sigma^n}=n(2-n)\gamma_{\sigma}+\frac{n(n-1)}{2}\gamma_{\sigma^2}
\end{equation}  
The combinatorics involved is simple: we first account for the anomalous dimension corresponding to $n$ propagators of $\sigma$, the second term then computes the propagator corrections for $\sigma^2$  (there are $n(n-1)/2$ such corrections). To avoid over-counting,  we subtract  $2\gamma_\sigma$ from each $\sigma^2$ correction, which ultimately gives the first term. 

\subsection{The $\lan \s^2 \s\ran$ correlator and conformal invariance}
At the short-range IR fixed point of the $O(N)$ vector model in $2<d<4$ dimensions
the composite operator $\hat\s^2$ is in general not a conformal primary \cite{Derkachov:1997gc,Derkachov:1998js}.
For instance, in general $d$ the correlator $\lan \hat\s^2\hat\s\ran$ is non-vanishing,
while the scaling dimensions of $\hat\s$ and $\hat\s^2$ are manifestly different.
However, the leading order scaling dimensions of the operators $\hat\s^2$
and $\partial^2 \hat\s$ are the same, and therefore due to the non-vanishing
$\lan \hat\s^2\hat\s\ran$ these operators can mix. In fact they do, and
result in a primary operator of the form $\hat\s^2 + \alpha \,\partial^2 \hat\s$ for a certain
$\alpha = {\cal O}(1/\sqrt{N})$ \cite{Derkachov:1997gc,Derkachov:1998js}.\footnote{Note, however,
that the mixing coefficient $\alpha$ vanishes for the integer-valued $d=2,3,4$.}

However, in the long-range CFT, due to the non-integer scaling dimension of $\s^2$, we do not expect it to mix with any other descendent operator. In fact, we expect it to be a conformal primary in the theory with $1\leq d<4$, and for any allowed value of $s$ in the long-range CFT region. One immediate manifestation of this fact should be borne out in the correlator
$\lan\s^2\s\ran$, which is expected to vanish. In this section, we perform this check explicitly,
working at the leading order in $1/N$.

To the leading order, the $\lan\s^2\s\ran$ correlator is determined by the kite diagram with a $\phi$ loop in the middle:
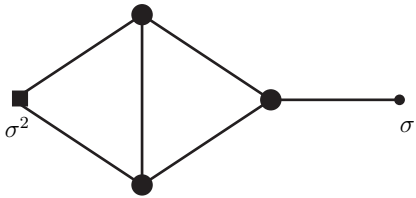
\begin{center}
  \begin{picture}(236,80) (27,5)
    \SetWidth{1.0}
    \SetColor{Black}
    \Line[](80,44)(128,76)
    \Line[](80,44)(128,12)
    \Line[](128,76)(128,12)
    \Line[](128,76)(176,44)
    \Line[](176,44)(128,12)
    \Line[](176,44)(224,44)
    \CBox(80,42)(85,47){Black}{Black}
    \Vertex(224,44){2}
    \Vertex(128,76){4}
    \Vertex(128,12){4}
    \Vertex(176,44){4}
    \Text(77,30)[lb]{\scalebox{0.8}{$\s^2$}}
    \Text(225,32)[lb]{\scalebox{0.8}{$\s$}}
  \end{picture}
\end{center}
One way to compute it is to replace the $\s^2\s\s$, and $\s\s\s$ vertices with their respective conformal triangles:\footnote{It should be noted that we introduce the conformal triangles to regulate the diagram even though it is finite.
However, the diagram is superficially divergent, and in order to calculate
it, we choose to re-write it in an equivalent form that possesses internal $\s$ lines.
The diagram is then regularized by a small shift $\delta$ of the exponents of internal
$\s$ propagators, even though the $\delta\rightarrow 0$ limit of such a diagram is finite.}
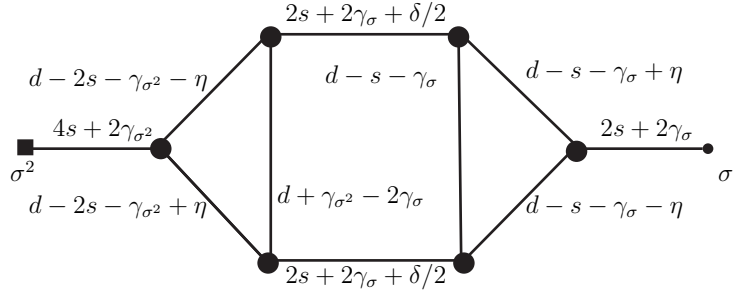
\begin{center}
  \begin{picture}(486,100) (-35,6)
    \SetWidth{1.0}
    \SetColor{Black}
    \Line[](61,57)(108,57)
    \Line[](109,57)(150,99)
    \Line[](152,100)(224,100)
    \Line[](111,57)(150,15)
    \Line[](151,15)(223,15)
    \Line[](152,100)(152,15)
    \Line[](222,100)(223,14)
    \Line[](223,100)(265,59)
    \Line[](265,57)(225,16)
    \Line[](266,57)(313,57)
    \CBox(58,55)(63,60){Black}{Black}
    \Vertex(315,57){2}
    \Text(55,45)[lb]{\scalebox{0.8}{$\s^2$}}
    \Text(320,45)[lb]{\scalebox{0.8}{$\s$}}
    \Line[](111,57)(150,15)
    \Vertex(111,57){4}
    \Vertex(152,99){4}
    \Vertex(151,14){4}
    \Vertex(222,99){4}
    \Vertex(224,14){4}
    \Vertex(266,56){4}
    \Text(71,60)[lb]{\scalebox{0.8}{$4s+2\gamma_{\s^2}$}}
    \Text(62,79)[lb]{\scalebox{0.8}{$d-2s-\gamma_{\s^2}-\eta$}}
    \Text(62,31)[lb]{\scalebox{0.8}{$d-2s-\gamma_{\s^2}+\eta$}}
    \Text(155,35)[lb]{\scalebox{0.8}{$d+\gamma_{\s^2}-2\gamma_\s$}}
    \Text(173,80)[lb]{\scalebox{0.8}{$d-s-\gamma_\s$}}
    \Text(158,103)[lb]{\scalebox{0.8}{$2s+2\gamma_\s+\delta/2$}}
    \Text(158,4)[lb]{\scalebox{0.8}{$2s+2\gamma_\s+\delta/2$}}
    \Text(248,82)[lb]{\scalebox{0.8}{$d-s-\gamma_{\s}+\eta$}}
    \Text(248,31)[lb]{\scalebox{0.8}{$d-s-\gamma_{\s}-\eta$}}
    \Text(276,60)[lb]{\scalebox{0.8}{$2s+2\gamma_{\s}$}}
  \end{picture}
\end{center}
Integrating over the internal vertices, while introducing auxiliary regulators wherever necessary
(such as choosing $\eta = \delta/2$ in the diagram above), we arrive at the final result given by
\begin{align}
 &\lan \s^2(x)\s(0)\ran = \f12(-6Z_{\s\s\s}^{(0)})(-2 Z_{\s^2\s\s}^{(0)})C_{\s^2}C_\s^3
 U\left(\f{d{-}\g_{\s^2}}2{-}s{-}\f{\d}4,s{+}\g_{\s}{+}\f{\d}4,\f{d{+}\g_{\s^2}}2{-}\g_\s\right)\nn\\
 &\times U\left(\f{d-s}2,\f{d-s}2-\f{\d}4,s+\f{\d}4\right)U\left(d-\f{3s}2,\f{3s}2+\f{\d}2,-\f{\d}2\right)U\left(s,\f{d+s+\d}2,\f{d-3s-\d}2\right)\nn\\
 &\times U\left(d-2s,\f{3s+\d}2,\f{s-\d}2\right)U\left(2s,\f{d-s+\d}2,\f{d-3s-\d}2\right)\,\f{\mu^{-\d}}{|x|^{3s+\d}}\,.
 \label{s2-s}
\end{align}
Here the large-$N$ amplitudes of the $\s^2\s\s$ and $\s\s\s$ conformal
triangles are given by\footnote{See section~\ref{sec:OPE coefficients}
for detailed discussion of conformal triangles in the long-range critical vector model.}
\begin{align}
 Z_{\s^2\s\s}^{(0)}=-\f12\f{C_{\s^2\s\s}^{(0)}}{C_{\s^2}C_{\s}^2 u_{\s^2\s\s}^{(0)}} = -\f1{C_{\s^2}u_{\s^2\s\s}^{(0)}},\hspace{1cm}Z_{\s\s\s}^{(0)} = \f{4C_\phi^3}3\,,
\end{align}
where we have used the leading-order amplitude of the $\langle \s^2\s\s\rangle$ three-point function
\begin{equation}
C_{\s^2\s\s}^{(0)} = 2C_{\s}^2\,,
\end{equation}
as well as the propagator amplitude (\ref{leading order s2 prop amplitude and dim}).
The expression for $u_{\s^2\s\s}^{(0)}$ is obtained by integrating over the vertices of the conformal triangle for $\s^2\s\s$, and then keeping only the leading large $N$ terms,\footnote{See \cite{Chai:2021uhv} for analogous
calculation in the short-range $O(N)$ vector model, where the corresponding $s^2 s s$
conformal triangle is calculated at the next-to-leading order in $1/N$ expansion.}
\begin{align}
 u_{\s^2\s\s}^{(0)} = \f2{2\gamma_\s- \gamma_{\s^2}}\frac{\pi ^{3 d/2} \Gamma \left(\frac{d}{2}-2 s\right) \Gamma \left(2
   s-\frac{d}{2}\right)}{
   \Gamma \left(\frac{d}{2}\right) \Gamma (2 s) \Gamma (d-2 s)}\,.
\end{align}
The factor of $2\gamma_\s -\gamma_{\s^2}$ is cancelled precisely by a leading large $N$ expansion of the only factor in \eqref{s2-s} whose dependence on the anomalous dimensions we retain for the sake of regularizing
the corresponding $U$ function, namely
\begin{align}
 U\left(\f{d-\g_{\s^2}}2-s-\f{\d}4,s+\g_{\s}+\f{\d}4,\f{d+\g_{\s^2}}2-\g_\s\right)=\frac{2\pi ^{d/2} }
 {\left(2\gamma_\s -\g_{\s^2} \right)\Gamma
   \left(\frac{d}{2}\right)} + \mo(N^0)\,.
\end{align}
Combining everything together and taking the limit $\delta\rightarrow 0$, we obtain
\begin{align}
 &\lan \s^2(x)\s(0)\ran = 0+\mo\left(\f1{N^{3/2}}\right)\,.
\end{align}
The vanishing of the leading order correlator between $\s$ and $\s^2$ provides further evidence that $\s^2$ is indeed a primary at the long-range fixed point, and that the fixed point in fact enjoys the full conformal symmetry. As an additional check, one can compute the first sub-leading order correction in the $1/N$ expansion, and show that it also vanishes for any $1\leq d<4$, and the corresponding allowed values of $s$ for the long-range CFT region.

\subsection{$\langle\sigma\phi\,\sigma\phi\rangle$}
\label{sec: sigma phi}

In this section we will discuss the operator $\sigma\phi$,
which will play an important role in section~\ref{sec: duality},
where we will develop the $O(N)$ generalization of the
short-range--long-range duality proposal of \cite{Behan:2017dwr}.
This composite operator has two constituents, one of which, $\sigma$,
possesses an anomalous dimension. However, interestingly enough, 
the composite operator $\sigma\phi$ itself does not acquire
anomalous dimension to all orders in $1/N$.
This exact result follows immediately from the $\phi$ e.o.m.
due to the action (\ref{eff-action-1}),
relating dimension of $\sigma\phi$ to the dimension of $\phi$:
%
\begin{equation}
\label{phi eom in LR}
\sigma\phi = -\sqrt{N}\,C(s)\,\int d^dy\,\frac{\phi(y)}{|x-y|^{d+s}}\,.
\end{equation}
Using the $\lan\phi\phi\ran$ two-point function (\ref{<phi phi> general}), while 
taking into account $\gamma_\phi = 0$, we obtain
\begin{equation}
\label{sigma phi propagator}
\lan \sigma\phi(x)\sigma\phi(0)\ran = N\,C(s)^2C_\phi\,(1+A_\phi)\,\pi^d A\left(
\frac{d-s}{2}\right) A\left(\frac{d+s}{2}\right)\,\frac{1}{|x|^{d+s}}\,,
\end{equation}
where we also took advantage of the identity (\ref{inverse propagator}).
We will be ignoring the $1/N$ corrections to the propagator amplitude, 
while noticing that the scaling dimension of the composite operator $\sigma\phi$ is given by 
\begin{equation}
\label{Delta sigma phi}
\Delta_{\sigma\phi} = \frac{d+s}{2}\,,
\end{equation}
exactly to all orders in $1/N$, despite the fact that the constituent $\sigma$ has
a non-trivial anomalous dimension.

As a consistency check, the relation (\ref{Delta sigma phi})
can be quickly verified at the next-to-leading order in $1/N$ expansion.
The first contribution is obtained by dressing of the internal
$\s$ and $\phi$ lines of the leading-order diagram:
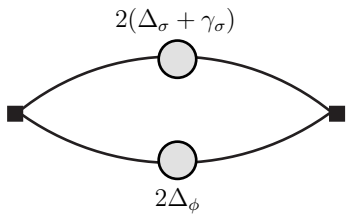
\begin{center}
  \begin{picture}(368,65) (-70,10)
    \SetWidth{1.0}
    \SetColor{Black}
    \CBox(45,35)(50,40){Black}{Black}
    \Arc[clock](107,-32.976)(91.978,129.496,50.504)
    \Arc[](107,117.5)(98.501,-127.161,-52.839)
    \GOval(108,58)(7,7)(0){0.882}
    \GOval(108,20)(7,7)(0){0.882}
    \CBox(165,35)(170,40){Black}{Black}
    \Text(85,68)[lb]{\scalebox{0.8}{$2(\Delta_\s+\gamma_\s)$}}
    \Text(100,0)[lb]{\scalebox{0.8}{$2\Delta_\phi$}}
  \end{picture}
\end{center}
Its contribution to the anomalous dimension is then given by
\begin{equation}
\label{gamma phi s 1}
\gamma_{\s\phi}^{(1)} = \gamma_\s\,.
\end{equation}
Next, we have the contribution due to the effective $\s\s\s$ vertex inserted into the
leading-order diagram:
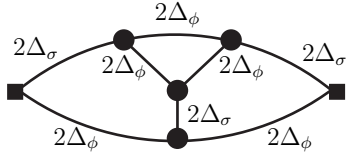
\begin{center}
  \begin{picture}(368,65) (-70,10)
    \SetWidth{1.0}
    \SetColor{Black}
    \CBox(45,35)(50,40){Black}{Black}
    \Arc[clock](107,-32.976)(91.978,129.496,50.504)
    \Arc[](107,117.5)(98.501,-127.161,-52.839)
    \Line[](108,20)(108,38)
    \Line[](108,38)(88,57)
    \Line[](108,38)(128,57)
    \Vertex(108,20){4}
    \Vertex(108,38){4}
    \Vertex(88,57){4}
    \Vertex(128,57){4}
    \CBox(165,35)(170,40){Black}{Black}
    \Text(100,62)[lb]{\scalebox{0.8}{$2\Delta_\phi$}}
    \Text(80,42)[lb]{\scalebox{0.8}{$2\Delta_\phi$}}
    \Text(124,42)[lb]{\scalebox{0.8}{$2\Delta_\phi$}}
    \Text(155,48)[lb]{\scalebox{0.8}{$2\Delta_\s$}}
    \Text(47,50)[lb]{\scalebox{0.8}{$2\Delta_\s$}}
    \Text(62,15)[lb]{\scalebox{0.8}{$2\Delta_\phi$}}
    \Text(142,15)[lb]{\scalebox{0.8}{$2\Delta_\phi$}}
    \Text(112,25)[lb]{\scalebox{0.8}{$2\Delta_\s$}}
  \end{picture}
\end{center}
This diagram is divergent. Since we are only interested in anomalous dimensions, we can regularize
it by simply extracting the logarithmically divergent term of the last integral.\footnote{This can be done by using the relation
\begin{equation}
\label{log regularization}
\int d^d y\,\frac{1}{|y|^d|x-y|^d} \supset 2\frac{2\pi^\frac{d}{2}}{\Gamma\left(\frac{d}{2}\right)}\,\log(\mu)\,\frac{1}{|x|^d}\,.
\end{equation}
}
As a result we obtain contribution to the anomalous dimension given by
\begin{equation}
\label{gamma phi s 2}
\gamma_{\sigma\phi}^{(2)} = -\frac{16}{N}C_\phi^4C_\s^2\frac{2\pi^\frac{d}{2}}{\Gamma\left(\frac{d}{2}\right)}
U\left(\frac{d-s}{2},\frac{d-s}{2},s\right)^2 U\left(s,\frac{s}{2},d-\frac{3s}{2}\right)\,.
\end{equation}
Finally, we have contribution from the following diagram:
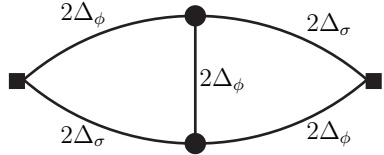
\begin{center}
  \begin{picture}(164,64) (31,20)
    \SetWidth{1.0}
    \SetColor{Black}
    \Arc[clock](104,-25.333)(97.333,131.112,48.888)
    \Arc[clock](104,121.333)(97.333,-48.888,-131.112)
    \CBox(35,45)(40,50){Black}{Black}
    \CBox(168,45)(173,50){Black}{Black}
    \Vertex(104,72){4}
    \Vertex(104,24){4}
    \Line[](104,24)(104,72)
    \Text(54,68)[lb]{\scalebox{0.8}{$2\Delta_\phi$}}
    \Text(54,23)[lb]{\scalebox{0.8}{$2\Delta_\s$}}
    \Text(146,65)[lb]{\scalebox{0.8}{$2\Delta_\s$}}
    \Text(146,23)[lb]{\scalebox{0.8}{$2\Delta_\phi$}}
    \Text(106,43)[lb]{\scalebox{0.8}{$2\Delta_\phi$}}
  \end{picture}
\end{center}
Regularizing it using (\ref{log regularization}) we obtain
\begin{equation}
\label{gamma phi s 3}
\gamma_{\sigma\phi}^{(3)} = -\frac{8}{N}C_\phi^2C_\s\frac{2\pi^\frac{d}{2}}{\Gamma\left(\frac{d}{2}\right)}
U\left(\frac{d-s}{2},\frac{d-s}{2},s\right)\,.
\end{equation}
Combining (\ref{gamma phi s 1}), (\ref{gamma phi s 2}), (\ref{gamma phi s 3}),
while taking into account $\gamma_\s$ given by (\ref{gamma_sigma result}),\footnote{Alternatively, this calculation
can be viewed as a consistency check for the value of $\gamma_\s$.}
we arrive at
\begin{equation}
\gamma_{\sigma\phi} = \gamma_{\sigma\phi}^{(1)}+\gamma_{\sigma\phi}^{(2)}+\gamma_{\sigma\phi}^{(3)}
=0+{\cal O}\left(\frac{1}{N^2}\right)\,,
\end{equation}
in agreement with the general argument given above.

\section{OPE coefficients}
\label{sec:OPE coefficients}

We now proceed to calculating various OPE 
coefficients in the long-range $O(N)$ critical vector model.
In section~\ref{sec: phi phi sigma} we calculate the 
$\langle \phi\phi \sigma\rangle$ three-point function, and determine
its amplitude at the next-to-leading order in $1/N$ expansion.
Such a calculation requires knowing the next-to-leading
order corrections to the $\phi\phi \sigma$ interaction vertex.
We calculate this non-local effective vertex using the background
field method \cite{Goykhman:2020ffn}. In section~\ref{sec: sigma sigma sigma} we calculate
the $\langle \sigma\sigma\sigma\rangle$
three-point function at the leading order in $1/N$ expansion.

\subsection{$\langle \phi\phi \sigma\rangle$}
\label{sec: phi phi sigma}

In this section we will calculate the $\langle \phi\phi \sigma\rangle$ correlation function
at the next-to-leading order in $1/N$ expansion.
Due to conformal symmetry, this three-point function has the form
\begin{equation}
\label{conformal form of <phi phi s>}
\langle \phi(x_1)\phi(x_2)\sigma(x_3)\rangle
=\frac{C_{\phi\phi\sigma}^{(0)}(1+\delta C_{\phi\phi\sigma})\mu^{-\gamma_\sigma}}{(|x_{13}||x_{23}|)^{\Delta_\sigma+\gamma_\sigma}
|x_{12}|^{2\Delta_\phi-\Delta_\sigma-\gamma_\sigma}}\,,
\end{equation}
where $\mu$ is an arbitrary mass scale.
The amplitude of the three-point function (\ref{conformal form of <phi phi s>}) 
was separated into the leading order factor of $C_{\phi\phi\sigma}^{(0)}$
and the relative $1/N$ corrections to it, $\delta C_{\phi\phi\sigma}={\cal O}(1/N)$.
We also took into account that the anomalous dimension of $\phi$
vanishes, and therefore its total dimension is given by $\Delta_\phi$.

Carrying out the calculation of (\ref{conformal form of <phi phi s>})
in this section, we will reproduce
the anomalous dimension $\gamma_\sigma$, and derive the leading order 
amplitude $C_{\phi\phi\sigma}^{(0)}$ and the next-to-leading order contribution to 
$\delta C_{\phi\phi\sigma}$. In the process, we will introduce and calculate the $\phi\phi \sigma$ conformal
triangle, at the next-to-leading order in $1/N$, representing the corresponding
non-local vertex in the effective action.

It is customary to use conventions in which position-space propagators of the fields in the considered
correlation function are normalized to unity. This can be achieved by the corresponding
rescaling of the fields,
\begin{equation}
\label{propagator unit normalization condition}
\phi\rightarrow \sqrt{C_\phi (1+A_\phi)}\,\phi\,,\qquad
\sigma\rightarrow \sqrt{C_\sigma (1+A_\sigma)}\,\sigma\,.
\end{equation}
In particular, such a normalization provides a renormalization scheme choice 
related to the freedom of redefining the scale $\mu$ by a constant factor.
We will denote the amplitudes of the correlation functions
with so-normalized fields with a bar, \textit{e.g.}, $\bar C_{\phi\phi\sigma}^{(0)}$
and $\delta \bar C_{\phi\phi\sigma}$ for the three-point function (\ref{conformal form of <phi phi s>})
in which all of the fields have been normalized according to (\ref{propagator unit normalization condition}).

The leading order amplitude $C_{\phi\phi s}^{(0)}$ can be easily found from the tree-level
diagram
\begin{center}
  \begin{picture}(500,100) (126,110)
    \SetWidth{1.0}
    \SetColor{Black}
    \Line[](217,159)(217,205)
    \Line[](217,159)(181,122)
    \Line[](250,122)(217,159)
    \Vertex(217,159){4}
    \Vertex(217,205){2}
    \Vertex(181,122){2}
    \Vertex(250,122){2}
    \Text(205,210)[lb]{\scalebox{1}{$\sigma (x_3)$}}
    \Text(150,105)[lb]{\scalebox{1}{$\phi (x_1)$}}
    \Text(255,105)[lb]{\scalebox{1}{$\phi (x_2)$}}
    \Text(285,150)[lb]{\scalebox{1}{$=-\frac{2}{\sqrt{N}}C_\phi^2 C_\sigma 
    U\left(\Delta_\phi,\Delta_\phi,\Delta_\sigma\right)
    \frac{1}{(|x_{13}||x_{23}|)^{\Delta_\sigma}|x_{12}|^{2\Delta_\phi-\Delta_\sigma}}$}}
   \end{picture}
\end{center}
For the fields normalized according to (\ref{propagator unit normalization condition}) we obtain
the leading-order amplitude
\begin{align}
\label{hat Cphiphi sigma}
\bar C_{\phi\phi\sigma}^{(0)} &= -\frac{2}{\sqrt{N}}C_\phi C_\sigma^\frac{1}{2} U\left(\Delta_\phi,\Delta_\phi,
\Delta_\sigma\right)\\
&=-\frac{1}{\sqrt{N}}\frac{\Gamma \left(\frac{s}{2}\right)^2 \Gamma \left(\frac{d}{2}-s\right) \sqrt{(d-2 s) \Gamma (s) \sin \left(\frac{1}{2} \pi  (d-2 s)\right) \Gamma (d-s)}}{\sqrt{\pi } \Gamma (s) \Gamma \left(\frac{d-s}{2}\right)^2}\,.\notag
\end{align}

To determine correction $\delta C_{\phi\phi\sigma}$ to the amplitude of the $\langle\phi\phi\sigma\rangle$
three-point function, we begin by calculating the corresponding $\phi\phi\sigma$ conformal triangle (see
\cite{Polyakov:1970xd} for the original discussion of conformal triangles in CFTs).
In terms of the effective action, such a conformal triangle gives a diagrammatic representation
of the non-local interaction term
\begin{equation}
S_{\textrm{eff}} \supset \frac{Z_{\phi\phi\sigma}}{\sqrt{N}}\,\mu^{\gamma_\sigma}\int d^dx_{1,2,3}\,
\frac{\phi(x_1)\phi(x_2)\sigma(x_3)}{(|x_{13}||x_{23}|)^{2\alpha}|x_{12}|^{2\beta}}\,,
\end{equation}
where we denoted
\begin{equation}
\alpha = \frac{d-s-\gamma_\sigma}{2}\,,\qquad\beta = s+\frac{\gamma_\sigma}{2}\,.
\end{equation}
The amplitude of the conformal triangle admits the $1/N$ expansion,
$Z_{\phi\phi\sigma}=Z_{\phi\phi\sigma}^{(0)}(1+\delta Z_{\phi\phi\sigma})$. Here the
leading order amplitude $Z_{\phi\phi\sigma}^{(0)} = {\cal O}(1)$ is constrained by the 
requirement that the classical interaction term $\frac{1}{\sqrt{N}}\phi^2\sigma$
is reproduced in the large-$N$ limit. Incorporating the next-to-leading order correction,
and taking into account the vertex counter-term contribution, we obtain the following
diagrammatic equation for the $\phi\phi \sigma$ conformal triangle:
\begin{center}
\begin{equation}
  \begin{picture}(500,193) (6,3)
    \SetWidth{1.0}
    \SetColor{Black}
    \Text(62,157)[lb]{\scalebox{0.6}{$2\alpha$}}
    \Text(100,157)[lb]{\scalebox{0.6}{$2\alpha$}}
    \Text(82,132)[lb]{\scalebox{0.6}{$2\beta$}}
    \Line[](85,203)(85,167)
    \Vertex(85,167){4}
    \Line[](85,167)(65,140)
    \Vertex(65,140){4}
    \Line[](104,141)(85,167)
    \Vertex(104,141){4}
    \Line[](66,141)(104,141)
    \Line[](65,140)(38,122)
    \Line[](130,122)(104,141)
    \Text(25,110)[lb]{\scalebox{1}{$\phi$}}
    \Text(133,110)[lb]{\scalebox{1}{$\phi$}}
    \Text(72,200)[lb]{\scalebox{1}{$\sigma$}}
    \Text(156,160)[lb]{\scalebox{1}{$=$}}
    \Line[](217,159)(217,205)
    \Line[](217,159)(181,122)
    \Line[](250,122)(217,159)
    \Vertex(217,159){4}
    \Text(205,200)[lb]{\scalebox{1}{$\sigma$}}
    \Text(168,110)[lb]{\scalebox{1}{$\phi$}}
    \Text(255,110)[lb]{\scalebox{1}{$\phi$}}
    \Text(285,160)[lb]{\scalebox{1}{$+$}}
    \Text(300,157)[lb]{\scalebox{1}{$\frac{\gamma_\sigma}{\delta}\times$}}
    \Line[](387,159)(387,205)
    \Line[](387,159)(352,122)
    \Line[](421,122)(387,159)
    \Vertex(387,159){4}
    \Text(375,200)[lb]{\scalebox{1}{$\sigma$}}
    \Text(338,110)[lb]{\scalebox{1}{$\phi$}}
    \Text(428,110)[lb]{\scalebox{1}{$\phi$}}
    \Text(80,42)[lb]{\scalebox{1}{$+\mu^{-\delta}\times$}}
    \Text(136,30)[lb]{\scalebox{0.6}{$2\Delta_\phi$}}
    \Text(188,30)[lb]{\scalebox{0.6}{$2\Delta_\phi$}}
    \Text(156,13)[lb]{\scalebox{0.6}{$2\Delta_\sigma+\delta$}}
    \Line[](167,88)(167,51)
    \Vertex(167,51){4}
    \Line[](167,51)(148,23)
    \Line[](187,23)(167,51)
    \Line[](148,23)(187,23)
    \Vertex(148,23){4}
    \Vertex(187,23){4}
    \Line[](148,23)(120,4)
    \Line[](213,4)(187,23)
    \Text(155,85)[lb]{\scalebox{1}{$\sigma$}}
    \Text(107,-5)[lb]{\scalebox{1}{$\phi$}}
    \Text(217,-5)[lb]{\scalebox{1}{$\phi$}}
    \Text(270,42)[lb]{\scalebox{1}{$+\mu^{-\delta}\times$}}
    \Text(310,30)[lb]{\scalebox{0.6}{$2\Delta_\sigma+\delta/2$}}
    \Text(393,30)[lb]{\scalebox{0.6}{$2\Delta_\sigma+\delta/2$}}
    \Text(362,13)[lb]{\scalebox{0.6}{$2\Delta_\phi$}}
    \Text(362,31)[lb]{\scalebox{0.6}{$2\Delta_\phi$}}
    \Text(342,57)[lb]{\scalebox{0.6}{$2\Delta_\phi$}}
    \Text(380,57)[lb]{\scalebox{0.6}{$2\Delta_\phi$}}
    \Line[](367,88)(367,73)
    \Vertex(367,73){4}
    \Line[](367,73)(340,23)
    \Line[](391,23)(367,73)
    \Line[](350,42)(382,42)
    \Vertex(350,42){4}
    \Vertex(382,42){4}
    \Line[](391,23)(340,23)
    \Vertex(391,23){4}
    \Vertex(340,23){4}
    \Line[](340,23)(322,4)
    \Line[](410,4)(391,23)
    \Text(312,-5)[lb]{\scalebox{1}{$\phi$}}
    \Text(410,-5)[lb]{\scalebox{1}{$\phi$}}
    \Text(355,85)[lb]{\scalebox{1}{$\sigma$}}
  \end{picture}
   \label{Dressed vertex diagram vm}
\end{equation}
\end{center}
The first term in the r.h.s. of (\ref{Dressed vertex diagram vm}) represents the tree-level
contribution of the leading-order $\phi^2\sigma$ interaction vertex,
while the second term stands for the counter-term contribution due to the
wave-function renormalization of the field $\sigma$.
The last two terms in the r.h.s. of (\ref{Dressed vertex diagram vm})
originate due to loop corrections to the $\phi^2\sigma$ vertex at the next-to-leading
order in $1/N$. In the latter diagrams, we have
adjusted the scaling dimension of the corresponding graphs by a small shift $\delta$
by regularizing the internal $\sigma$ lines.
At the end of the calculation we will take the limit $\delta\rightarrow 0$.
We will also observe explicitly that the total $1/\delta$ pole of the second and third
diagrams is cancelled out precisely by the second (counter-term) diagram.
Inversely, imposing a cancellation of divergencies reproduces the correct value for the anomalous
dimension $\gamma_\sigma$ given by (\ref{gamma_sigma result}).

The $\phi\phi\sigma$ conformal triangle
can be used directly to calculate correlation functions and extract CFT data.
For instance, in the previous paragraph, we outlined how the anomalous dimension $\gamma_\sigma$ can be found
from such a calculation.
Importantly, OPE coefficients can be calculated as well.
To this end, one needs
to attach to the conformal triangle the full (dressed) propagators, and integrate over the internal
unique vertices: such a procedure is a direct generalization of using the Feynman rule
for a tree-level vertex and integrating over an insertion of the vertex. In particular, to calculate the
$\langle\phi\phi\sigma\rangle$ three-point function we proceed as follows:
\begin{center}
\begin{equation}
  \begin{picture}(500,93) (-6,103)
    \SetWidth{1.0}
    \SetColor{Black}
    \Text(62,157)[lb]{\scalebox{0.6}{$2\alpha$}}
    \Text(100,157)[lb]{\scalebox{0.6}{$2\alpha$}}
    \Text(82,132)[lb]{\scalebox{0.6}{$2\beta$}}
    \Text(92,184)[lb]{\scalebox{0.6}{$2(\Delta_\sigma+\gamma_\sigma)$}}
    \Text(122,130)[lb]{\scalebox{0.6}{$2\Delta_\phi$}}
    \Text(32,130)[lb]{\scalebox{0.6}{$2\Delta_\phi$}}
    \Line[](85,203)(85,167)
    \Vertex(85,167){4}
    \Vertex(85,203){2}
    \Line[](85,167)(65,140)
    \Vertex(65,140){4}
    \Line[](104,141)(85,167)
    \Vertex(104,141){4}
    \Vertex(38,122){2}
    \Vertex(130,122){2}
    \Line[](66,141)(104,141)
    \Line[](65,140)(38,122)
    \Line[](130,122)(104,141)
    \Text(20,105)[lb]{\scalebox{1}{$\phi (x_1)$}}
    \Text(133,105)[lb]{\scalebox{1}{$\phi (x_2)$}}
    \Text(52,200)[lb]{\scalebox{1}{$\sigma(x_3)$}}
    \Text(180,150)[lb]{\scalebox{1.2}{$=-\frac{\frac{2Z_{\phi\phi\sigma}}{\sqrt{N}}\mu^{-\gamma_\sigma}(C_\phi(1+A_\phi))^2
    C_\sigma(1+A_\sigma){\cal U}_{\phi\phi\sigma}}{(|x_{13}||x_{23}|)^{\Delta_\sigma+\gamma_\sigma}
|x_{12}|^{2\Delta_\phi-\Delta_\sigma-\gamma_\sigma}}$}}
  \end{picture}
     \label{<phi phi s> from conformal triangle}
\end{equation}
\end{center}
where we denoted the factor obtained from integration over three unique vertices of the
conformal triangle as
\begin{align}
{\cal U}_{\phi\phi\sigma}&=U\left(\frac{d-s-\gamma_\sigma}{2},\frac{d-s-\gamma_\sigma}{2},
s+\gamma_\sigma\right)
U\left(\frac{d-s}{2},\frac{d-\gamma_\sigma}{2},
\frac{s+\gamma_\sigma}{2}\right)\notag\\
&\times U\left(\frac{d-s}{2},\frac{d-s-\gamma_\sigma}{2},
s+\frac{\gamma_\sigma}{2}\right)\,.\label{U phi phi sigma}
\end{align}
Expanding (\ref{<phi phi s> from conformal triangle}) in $1/N$
to the next-to-leading order, and comparing the result with (\ref{conformal form of <phi phi s>})
we obtain
\begin{align}
\label{Zphiphis 0}
\bar C_{\phi\phi\sigma}^{(0)} &= -\frac{2Z_{\phi\phi\sigma}^{(0)}}{\sqrt{N}} \,C_\phi \, C_\sigma^\frac{1}{2}\,
{\cal U}_{\phi\phi\sigma}^{(0)} \,,\\
\delta\bar C_{\phi\phi\sigma}^{(0)} &= \delta Z_{\phi\phi\sigma} + \delta {\cal U}_{\phi\phi\sigma}
+A_\phi +\frac{A_\sigma}{2}\,.\label{delta C phiphi s in terms of Z}
\end{align}
where we denoted the $1/N$ expansion of (\ref{U phi phi sigma}) as
\begin{align}
{\cal U}_{\phi\phi\sigma}&={\cal U}_{\phi\phi\sigma}^{(0)}(1+\delta {\cal U}_{\phi\phi\sigma})\,,\notag\\
{\cal U}_{\phi\phi\sigma}^{(0)}&=\frac{2 \pi ^{\frac{3 d}{2}}  \Gamma \left(\frac{s}{2}\right)^4 \Gamma \left(\frac{d}{2}-s\right)^2}{\Gamma \left(\frac{d}{2}\right) \Gamma (s)^2 \Gamma \left(\frac{d-s}{2}\right)^4\,\gamma_\sigma}\,,\\
\delta {\cal U}_{\phi\phi\sigma}&{=}\frac{\gamma_\sigma}{2}\left(2 \psi ^{(0)}\left(\frac{d{-}s}{2}\right){-}3 \psi ^{(0)}\left(\frac{d}{2}{-}s\right){+}\psi ^{(0)}\left(\frac{d}{2}\right){+}2 \psi ^{(0)}\left(\frac{s}{2}\right){-}3 \psi ^{(0)}(s){-}\gamma\right)\,.\notag
\end{align}
Using (\ref{hat Cphiphi sigma}), (\ref{Zphiphis 0})
we can solve for the leading-order amplitude of the conformal triangle
\begin{equation}
\label{result for Zphiphis 0}
Z_{\phi\phi\sigma}^{(0)}=\frac{U\left(\Delta_\phi,\Delta_\phi,
\Delta_\sigma\right)}{{\cal U}_{\phi\phi\sigma}^{(0)}}
=\frac{ \Gamma \left(\frac{d}{2}\right) \Gamma (s) \Gamma \left(\frac{d-s}{2}\right)^2\, \gamma_\sigma }{2 \pi ^{d}\,\Gamma \left(\frac{s}{2}\right)^2 \Gamma \left(\frac{d}{2}-s\right)}\,.
\end{equation}
At the same time, (\ref{delta C phiphi s in terms of Z}) gives a prescription to calculate the $1/N$ correction to the OPE coefficient. To finish that calculation
we need to determine first the next-to-leading correction $\delta Z_{\phi\phi\sigma}$ 
to the $\phi\phi\sigma$ conformal triangle.

In \cite{Goykhman:2020ffn} it was proposed to use the background field method to calculate conformal triangles.
Such a method can be applied for the purpose of determining the values of $Z_{\phi\phi\sigma}^{(0)}$,
$\delta Z_{\phi\phi\sigma}$ in our case as well. Following \cite{Goykhman:2020ffn}, we set the field $\sigma$ to
a non-dynamical background value $\sigma\equiv \bar\sigma$, and attach the
full $\phi$ propagators to each  term on both sides of (\ref{Dressed vertex diagram vm}).
The resulting diagrammatic equation for the propagator $\langle \phi\phi\rangle |_{\bar \sigma}$
in the $\bar\sigma$ background is given by
\begin{center}
\begin{equation}
  \begin{picture}(500,150) (26,30)
    \SetWidth{1.0}
    \SetColor{Black}
     \Text(83,177)[lb]{\scalebox{1}{$\bar \sigma$}}
     \Text(236,170)[lb]{\scalebox{1}{$\bar \sigma$}}
     \Text(387,170)[lb]{\scalebox{1}{$\bar \sigma$}}
     \Text(146,70)[lb]{\scalebox{1}{$\bar \sigma$}}
     \Text(366,92)[lb]{\scalebox{1}{$\bar \sigma$}}
    \Text(62,157)[lb]{\scalebox{0.6}{$2\alpha$}}
    \Text(100,157)[lb]{\scalebox{0.6}{$2\alpha$}}
    \Text(82,132)[lb]{\scalebox{0.6}{$2\beta$}}
    \Vertex(85,167){4}
    \Line[](85,167)(65,140)
    \Vertex(65,140){4}
    \Line[](104,141)(85,167)
    \Vertex(104,141){4}
    \Line[](66,141)(104,141)
    \Line[](65,140)(38,122)
    \Line[](130,122)(104,141)
    \Vertex(38,122){2}
    \Vertex(130,123){2}
    \Text(30,130)[lb]{\scalebox{0.6}{$2\Delta_\phi$}} 
    \Text(123,130)[lb]{\scalebox{0.6}{$2\Delta_\phi$}} 
    \Text(163,145)[lb]{\scalebox{1}{$=$}}
    \Line[](237,159)(201,122)
    \Line[](270,122)(237,159)
    \Vertex(237,159){4}
    \Vertex(201,122){2}
    \Vertex(270,122){2}
    \Text(195,130)[lb]{\scalebox{0.6}{$2\Delta_\phi$}} 
    \Text(267,130)[lb]{\scalebox{0.6}{$2\Delta_\phi$}} 

    \Text(310,142)[lb]{\scalebox{1}{$+\frac{\gamma_\sigma}{\delta}\times$}}
    \Line[](387,159)(352,122)
    \Line[](421,122)(387,159)
    \Vertex(352,122){2}
    \Vertex(421,122){2}
    \Vertex(387,159){4}
    \Text(345,130)[lb]{\scalebox{0.6}{$2\Delta_\phi$}} 
    \Text(418,130)[lb]{\scalebox{0.6}{$2\Delta_\phi$}} 
    \Text(51,37)[lb]{\scalebox{1}{$+\mu^{-\delta}\times$}}
    \Text(116,40)[lb]{\scalebox{0.6}{$2\Delta_\phi$}}
    \Text(168,40)[lb]{\scalebox{0.6}{$2\Delta_\phi$}}
    \Text(136,23)[lb]{\scalebox{0.6}{$2\Delta_\sigma+\delta$}}
    \Vertex(147,61){4}
    \Line[](147,61)(128,33)
    \Line[](167,33)(147,61)
    \Line[](128,33)(167,33)
    \Vertex(128,33){4}
    \Vertex(167,33){4}
    \Line[](128,33)(100,14)
    \Line[](193,14)(167,33)
    \Vertex(100,14){2}
    \Vertex(193,14){2}
    \Text(92,20)[lb]{\scalebox{0.6}{$2\Delta_\phi$}} 
    \Text(190,20)[lb]{\scalebox{0.6}{$2\Delta_\phi$}} 
    \Text(245,37)[lb]{\scalebox{1}{$+\mu^{-\delta}\times$}}
    \Text(310,40)[lb]{\scalebox{0.6}{$2\Delta_\sigma+\delta/2$}}
    \Text(393,40)[lb]{\scalebox{0.6}{$2\Delta_\sigma+\delta/2$}}
    \Text(362,20)[lb]{\scalebox{0.6}{$2\Delta_\phi$}}
    \Text(362,41)[lb]{\scalebox{0.6}{$2\Delta_\phi$}}
    \Text(342,67)[lb]{\scalebox{0.6}{$2\Delta_\phi$}}
    \Text(380,67)[lb]{\scalebox{0.6}{$2\Delta_\phi$}}
    \Vertex(367,83){4}
    \Line[](367,83)(340,33)
    \Line[](391,33)(367,83)
    \Line[](350,52)(382,52)
    \Vertex(350,52){4}
    \Vertex(382,52){4}
    \Line[](391,33)(340,33)
    \Vertex(391,33){4}
    \Vertex(340,33){4}
    \Line[](340,33)(322,14)
    \Line[](410,14)(391,33)
    \Vertex(322,14){2}
    \Vertex(410,14){2}
    \Text(312,20)[lb]{\scalebox{0.6}{$2\Delta_\phi$}} 
    \Text(412,20)[lb]{\scalebox{0.6}{$2\Delta_\phi$}} 
  \end{picture}
   \label{Dressed scalar propagator in s background vm}
\end{equation}
\end{center}
Every diagram in (\ref{Dressed scalar propagator in s background vm})
can be readily calculated simply by applying the propagator merging relation,
while taking the integrals starting from the topmost vertex. In particular, the l.h.s. of 
(\ref{Dressed scalar propagator in s background vm}) gives
\begin{align}
\label{lhs of conformal triangle equation}
\textrm{l.h.s. \;\; of\;\;\;\;} (\ref{Dressed scalar propagator in s background vm}) \textrm{\;\;\;\;=\;\;}
-\frac{2}{\sqrt{N}}\, Z_{\phi\phi\sigma} \, (C_\phi(1+A_\phi))^2\,{\cal U}_{\phi\phi\sigma}\,\frac{\bar\sigma\,\mu^{\gamma_\sigma}}
{|x_{12}|^{d-2s-\gamma_\sigma}}\,,
\end{align}
while we denote the four terms contributing on the r.h.s. of (\ref{Dressed scalar propagator in s background vm})
 as
\begin{equation}
\label{rhs of conformal triangle equation}
\textrm{r.h.s. \;\; of\;\;\;\;} (\ref{Dressed scalar propagator in s background vm}) \textrm{\;\;\;\;=\;\;}
v_0+v_{\textrm{c.t.}} + v_1+v_2\,.
\end{equation}
Here the tree level diagram and the vertex counter-term contribute
\begin{align}
\label{v0}
v_0&=-\frac{2}{\sqrt{N}}\, (C_\phi(1+A_\phi))^2\,U\left(\frac{d-s}{2},\frac{d-s}{2},s\right)\,
\frac{\bar\sigma}{|x_{12}|^{d-2s}}\,,\\
v_{\textrm{c.t.}}&=-\frac{2}{\sqrt{N}}\, (C_\phi(1+A_\phi))^2
\,U\left(\frac{d-s}{2},\frac{d-s}{2},s\right)\frac{\gamma_\sigma}{\delta}\,\frac{\bar\sigma\,\mu^{\gamma_\sigma}}
{|x_{12}|^{d-2s-\gamma_\sigma}}\,.
\label{vct}
\end{align}
At the same time, for the vertex correction diagrams
in the second line of (\ref{Dressed scalar propagator in s background vm})
we obtain
\begin{align}
\label{v1}
v_1&=\left(-\frac{2}{\sqrt{N}}\right)^3C_\phi^4C_\sigma (1+A_\phi)^2
U\left(\frac{d-s}{2},\frac{d-s}{2},s\right)U\left(\frac{d-s}{2},\frac{d+\delta}{2},\frac{s-\delta}{2}\right)\notag\\
&\times U\left(\frac{d-s}{2},\frac{d-s+\delta}{2},s-\frac{\delta}{2}\right)
\,\frac{\bar\sigma\,\mu^{\gamma_\sigma}}{|x_{12}|^{d-2s+\delta}}\,,\\
\label{v2}
v_2&=4C_\phi^2C_\sigma 
U\left(d{-}\frac{3s}{2},s{+}\frac{\delta}{4},\frac{s}{2}{-}\frac{\delta}{4}\right)
U\left(s{+}\frac{\delta}{4},\frac{d{-}s}{2}{+}\frac{\delta}{4},\frac{d{-}s{-}\delta}{2}\right)\,v_1\,.
\end{align}
Comparing (\ref{lhs of conformal triangle equation}), (\ref{rhs of conformal triangle equation})
we first of all cancel the common factor of $-\frac{2}{\sqrt{N}}\, (C_\phi(1+A_\phi))^2\,\bar\sigma$.
Matching the leading order terms, we reproduce $Z_{\phi\phi \sigma}^{(0)}$
given by (\ref{result for Zphiphis 0}). Expanding the sub-leading contributions (\ref{v1}), (\ref{v2})
around $\delta=0$ we observe that the $1/\delta$ poles are exactly cancelled out by the
counter-term (\ref{vct}) for $\gamma_\sigma$ given by (\ref{gamma_sigma result}). At the same time, the anomalous dimension $\log |x_{12}|$ term 
in (\ref{v1}), (\ref{v2}) matches its counterpart on the l.h.s. (\ref{lhs of conformal triangle equation})
of the conformal triangle equation. Finally, matching the finite terms on both
sides of this equation, we obtain the next-to-leading order correction to the conformal triangle amplitude
\begin{align}
\delta Z_{\phi\phi \sigma}  &=-\frac{1}{N}\,\frac{4 \Gamma \left(\frac{s}{2}\right)^2 \Gamma (d-s)}
{\Gamma \left(\frac{d}{2}\right) \Gamma (s) \Gamma \left(d-\frac{3 s}{2}\right) \Gamma \left(\frac{d-s}{2}\right)^3 \Gamma \left(s-\frac{d}{2}\right)^2}\\
&\times \left(\Gamma (s) \Gamma \left(d{-}\frac{3 s}{2}\right) \Gamma \left(\frac{d{-}s}{2}\right) \Gamma \left(s{-}\frac{d}{2}\right){-}3 \Gamma \left(\frac{s}{2}\right) \Gamma \left(\frac{d}{2}{-}s\right) \Gamma (d{-}s) \Gamma \left(\frac{3 s}{2}{-}\frac{d}{2}\right)\right)
\notag\\
&\times\left( -\psi ^{(0)}\left(\frac{d-s}{2}\right)+\psi ^{(0)}\left(\frac{d}{2}-s\right)-\psi ^{(0)}\left(\frac{s}{2}\right)+\psi ^{(0)}(s)\right)+{\cal O}\left(\frac{1}{N^2}\right)\,.\notag
\end{align}

With all the ingredients in place, we can calculate the $1/N$
correction to the amplitude of the $\langle\phi\phi\sigma\rangle$
three-point function (\ref{delta C phiphi s in terms of Z}).
Our result for $\delta C_{\phi\phi\sigma}$ satisfies several consistency
checks. When $s=\frac{d}{2}$ the long-range fixed point becomes free,
and consequently $\delta C_{\phi\phi\sigma}|_{s=d/2}=0$,
as can be established by substituting $\sigma\sim \phi^2$ and performing Wick contractions.
This agrees with the $s\rightarrow d/2$ limit of (\ref{delta C phiphi s in terms of Z}).

When $s=1$, and $2<d<4$, the UV fixed point of the  long-range $O(N)$ vector model
was argued in \cite{Giombi:2019enr} to be critically equivalent to the IR fixed point
of an interacting `mixed $\sigma\phi$ theory' with local kinetic terms for $\phi$ and $\sigma$, cubic interaction $\phi^2\sigma$
and quartic interaction $\sigma^4$. The critical coupling in the latter model, calculated at the leading order in the $\epsilon$-expansion
performed around $d=4$ dimensions, gives \cite{Giombi:2019enr}
\begin{equation}
g_1^\star = 8\pi\,\sqrt{\frac{2\epsilon}{N-32}} + {\cal O}(\epsilon^{3/2})\,.
\end{equation}
This implies that in the `mixed $\s\phi$' theory,
\begin{equation}
\delta \bar C_{\phi\phi\sigma} |_{s=1,d=4-\epsilon} = \frac{16}{N} + {\cal O}(\epsilon^{3/2},1/N^2)\,,
\end{equation}
which agrees with (\ref{delta C phiphi s in terms of Z}) for $s=1$ expanded in $d=4-\epsilon$ dimensions.

\subsection{$\langle \sigma\sigma\sigma\rangle$}
\label{sec: sigma sigma sigma}

In this section we will calculate the $\langle \sigma\sigma\sigma\rangle$
three-point function at the leading order in the $1/N$ expansion.
Expanding the trace log in the large-$N$ effective action for $\sigma$ \eqref{eff-action-sigma} to ${\cal O}(1/N^{3/2})$ we obtain the following cubic term
\begin{equation}
S_{\textrm{eff}} \supset -\frac{1}{3!}\left(-\frac{2}{\sqrt{N}}\right)^3\,C_\phi^3\int d^dx_{1,2,3}
\frac{\sigma(x_1)\sigma(x_2)\sigma(x_3)}{(|x_{12}||x_{13}||x_{23})^{d-s}}\,.
\end{equation}
Such a non-local cubic interaction vertex can be
represented diagrammatically using the $\sigma\sigma\sigma$ conformal-triangle.
The corresponding Feynman rule is then given by
\begin{center}
  \begin{picture}(500,93) (-6,103)
    \SetWidth{1.0}
    \SetColor{Black}
    \Text(55,157)[lb]{\scalebox{0.6}{$d-s$}}
    \Text(100,157)[lb]{\scalebox{0.6}{$d-s$}}
    \Text(80,132)[lb]{\scalebox{0.6}{$d-s$}}
    \Line[](85,203)(85,167)
    \Vertex(85,167){4}
    \Line[](85,167)(65,140)
    \Vertex(65,140){4}
    \Line[](104,141)(85,167)
    \Vertex(104,141){4}
    \Line[](66,141)(104,141)
    \Line[](65,140)(38,122)
    \Line[](130,122)(104,141)
    \Text(20,105)[lb]{\scalebox{1}{$\sigma (x_1)$}}
    \Text(133,105)[lb]{\scalebox{1}{$\sigma (x_2)$}}
    \Text(52,200)[lb]{\scalebox{1}{$\sigma(x_3)$}}
    \Text(180,150)[lb]{\scalebox{1.2}{$=-\frac{6Z_{\sigma\sigma\sigma}}{\sqrt{N}}\,\frac{\sigma(x_1)\sigma(x_2)\sigma(x_3)}{(|x_{12}||x_{13}||x_{23})^{d-s}}$}}
  \end{picture}
\end{center}
Here the leading order amplitude of the conformal triangle is given by
\begin{equation}
Z_{\sigma\sigma\sigma}^{(0)} = \frac{4C_\phi^3}{3}\,.
\end{equation}
Attaching $\sigma$ propagators to the conformal triangle and integrating over three unique vertices
we obtain the three-point function for normalized $\sigma$,
\begin{equation}
\langle \sigma(x_1)\sigma(x_2)\sigma(x_3)\rangle\Bigg|_{\textrm{normalized}}
=\frac{\bar C_{\sigma\sigma\sigma}^{(0)}}{(|x_{12}||x_{13}||x_{23}|)^s}\,,
\end{equation}
where we denoted the normalized amplitude as
\begin{align}
\bar C_{\sigma\sigma\sigma}^{(0)}&=-\frac{8}{\sqrt{N}}C_\phi^3C_\sigma^\frac{3}{2}U\left(\frac{d-s}{2},\frac{d-s}{2},s\right)^2
U\left(s,\frac{s}{2},d-\frac{3s}{2}\right)\\
&=\frac{2^{d-2 s} (d-2 s) \sin \left(\frac{1}{2} \pi  (d-2 s)\right) \Gamma \left(\frac{d}{2}-s\right)^2 \Gamma \left(\frac{1}{2} (d-s+1)\right) \Gamma \left(\frac{3 s}{2}-\frac{d}{2}\right)}{\pi  \Gamma \left(\frac{s+1}{2}\right) \Gamma \left(d-\frac{3 s}{2}\right)}
\hat C_{\phi\phi\sigma}^{(0)}\,.\notag
\end{align}
A simple consistency check of this result is $\bar C_{\sigma\sigma\sigma}^{(0)}|_{s=1,
d=4-\epsilon} = 0+{\cal O}(\epsilon^\frac{3}{2})$, in agreement with the behavior of its counterpart
in the dual model \cite{Giombi:2019enr}.

\section{Continuity of CFT data across long-range--short-range crossover}
\label{sec: crossover to short range}

In previous sections we derived various CFT data for the long-range conformal fixed point
for a general exponent $s$ and space-time dimension $d$. At the same time, we know that
when $s=2$ the long-range CFT is classically equivalent to the short-range CFT; this statement is
evident by matching the actions of these two models (for instance, in momentum space). Therefore one might wonder if
taking the limit $s\rightarrow 2$ in the expressions for the long-range CFT data obtained
above, one would recover the known CFT data of the short-range $O(N)$ vector model.
As reviewed in Introduction, it is well known that such a naive check would not fulfill that expectation;
and the crossover between short-range and long-range CFTs in fact happens at the value $s=s_\star<2$ \cite{Sak1973,PhysRevB154344}. Here $s_\star=2-2\g_{\hat\phi}$, where $\g_{\hat\phi}$ is the  anomalous dimension of $\hat\phi$ in the short-range model. The value of $s_\star$ is defined by requiring a continuity of
CFT data
across the long-range to the short-range transition.\footnote{Note that the definition of the crossover point is independent of the value of $N$ (while the value of $s_\star$ is dependent on $N$), and can be easily generalized to the $O(N)$ model.} In this subsection, we provide explicit evidence for the existence
of a continuous transition of all CFT data obtained above
using the $1/N$ expansion, \footnote{Analogous  calculation in the perturbative $\epsilon$-expansion,
for $\epsilon = 2s-d$, was first carried out in \cite{Sak1973} to order $\epsilon^2$.} namely the scaling dimensions of $\sigma$, $\sigma^2$, $\sigma\phi$, and the OPE coefficients $C_{\phi\phi\sigma}$, and
$C_{\sigma\sigma\sigma}$, at the crossover point $s=s_\star$. This is a non-trivial consistency check of our long-range CFT data computed near a strongly-coupled point $s=s_\star$ in general $d$,  and at the next-to-leading order in $1/N$.

\subsection{Continuity of dimensions of $\sigma$, $\sigma^2$, and $\sigma\phi$}
\label{sec: continuity of dimensions}

Before proceeding onto the calculations in this section, we make the following simple observations regarding the leading order behaviour of the scaling dimensions and amplitudes of $\phi$ and $\s$,

\begin{align}
\label{s->2 simple limits}
\begin{split}
\Delta_\phi |_{s\rightarrow 2}&= \Delta_{\hat\phi}=\frac{d}{2}-1\,,\qquad
\Delta_\sigma |_{s\rightarrow 2}=\Delta_{\hat\sigma} =2\,,\\
C_\phi |_{s\rightarrow 2}&= C_{\hat\phi} =\frac{\Gamma\left(\frac{d}{2}-1\right)}
{4\pi^\frac{d}{2}}\,,\;\;
C_\sigma |_{s\rightarrow 2}= C_{\hat\sigma} =\frac{2^d\Gamma\left(\frac{d-1}{2}\right)
\sin\left(\frac{\pi d}{2}\right)}{\pi^\frac{3}{2}\Gamma\left(\frac{d}{2}-2\right)}\,,
\end{split}
\end{align}

To illustrate the continuity of scaling dimension of the Hubbard-Stratonovich field $\s$ across the crossover point $s_\star$
we will utilize the derivation of the anomalous dimension $\gamma_\sigma$
reviewed in section~\ref{sec: <ss>}.
The total scaling dimension of $\sigma$ is given by $\Delta_\sigma +\gamma_\sigma$, where
$\Delta_\sigma = s$. Notice that at the crossover point $s_\star=2-2\gamma_{\hat\phi}$,
$\Delta_\sigma$ gets split into
the sum of two parts, $\Delta_{\hat\sigma}=2$, that equals to the leading order contribution
to scaling dimension of $\hat\s$, and the sub-leading
term $-2\gamma_{\hat\phi}$. For future purposes, let us write down this rearrangement of total scaling dimension 
of $\sigma$ in the long-range CFT at $s_\star$ as
\begin{equation}
(\Delta_\sigma +\gamma_\sigma)|_{s\rightarrow s_\star}
= \Delta_{\hat\sigma} +(\gamma_\sigma|_{s\rightarrow s_\star} - 2\gamma_{\hat\phi})\,.
\label{Delta sigma at crossover}
\end{equation}
Notice that terms in brackets in r.h.s. of (\ref{Delta sigma at crossover})
are sub-leading in $1/N$, and therefore taking the limit $s\rightarrow s_\star = 2+{\cal O}(1/N)$
in those terms (more precisely, in $\gamma_\sigma$, since $\gamma_{\hat\phi}$
does not depend on $s$) needs to be replaced with taking the limit $s\rightarrow 2$.
In order for the scaling dimension of the $\s$ field
to be continuous at the crossover point $s_\star$,
\begin{equation}
(\Delta_\sigma+\gamma_\sigma)|_{s\rightarrow s_\star} =  \Delta_{\hat\sigma} + \gamma_{\hat\sigma}\,,
\end{equation}
it is then required that
\begin{equation}
\label{s->2 limit of gamma sigma}
\gamma_\sigma|_{s\rightarrow 2} = \gamma_{\hat\sigma}+2\gamma_{\hat\phi}\,.
\end{equation}

To verify that the above expression is true,
we notice that
 among the three diagrams contributing to $\langle\sigma\sigma\rangle$
at the next-to-leading order in $1/N$ (discussed in section~\ref{sec: <ss>}),
two of them, namely $C^{(1,2)}_{\sigma\sigma}$, reproduce the corresponding diagrams
in the short-range critical vector model when $s\rightarrow 2$, i.e. $C^{(1,2)}_{\sigma\sigma}|_{s\rightarrow 2}= C^{(1,2)}_{\hat\sigma\hat\sigma}$.
However, the third diagram,
$C^{(3)}_{\sigma\sigma}$,
does not reduce to its short-range counterpart $C^{(3)}_{\hat\sigma\hat\sigma}$ in that limit.
Its contribution to $\gamma_\sigma$ is in fact absent, unlike the
contribution of $C^{(3)}_{\hat\sigma\hat\sigma}$ to $\gamma_{\hat\sigma}$.
To find the latter, we consider $C^{(3)}_{\hat\sigma\hat\sigma}$ diagram explicitly:\footnote{See \cite{Goykhman:2019kcj} for a recent detailed calculation of anomalous dimension of the Hubbard-Stratonovich field.}
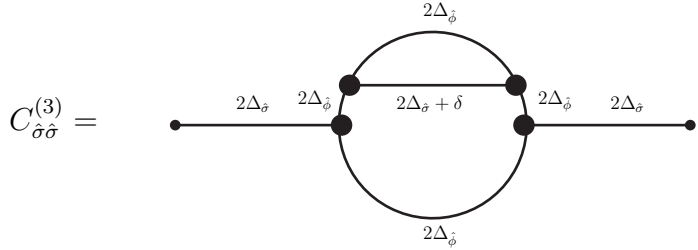
\begin{center}
  \begin{picture}(194,90) (31,-27)
    \SetWidth{1.0}
    \SetColor{Black}
    \Text(-30,5)[lb]{$ C^{(3)}_{\hat\sigma\hat\sigma}=$}
    \Line[](32,10)(94,10)
    \Text(55,15)[lb]{\scalebox{0.6}{$2\Delta_{\hat\s}$}}
    \Text(78,15)[lb]{\scalebox{0.6}{$2\Delta_{\hat\phi}$}}
    \Text(168,15)[lb]{\scalebox{0.6}{$2\Delta_{\hat\phi}$}}
    \Arc[](128,10)(35,153,513)
    \Line[](162,10)(224,10)
    \Text(195,15)[lb]{\scalebox{0.6}{$2\Delta_{\hat\s}$}}
    \Line[](97,25)(159,25)
    \Text(115,15)[lb]{\scalebox{0.6}{$2\Delta_{\hat\s}+\delta$}}
    \Text(125,48)[lb]{\scalebox{0.6}{$2\Delta_{\hat\phi}$}}
    \Text(125,-36)[lb]{\scalebox{0.6}{$2\Delta_{\hat\phi}$}}
    \Vertex(32,10){2}
    \Vertex(94,10){4}
    \Vertex(162,10){4}
    \Vertex(224,10){2}
    \Vertex(97,25){4}
    \Vertex(159,25){4}
  \end{picture}
\end{center}
As pointed out above, this diagram is divergent
in the short-range model, but its counterpart in the long-range model is finite.
Importantly, the divergent behavior of this graph can be traced back to its sub-diagram,
representing the
$1/N$ correction to the propagator of the field $\hat\phi$. However, as shown
in section~\ref{sec: phi phi correction}, the one-loop correction to the propagator of $\phi$ is finite in the long-range CFT, consistently with the
expectation that scaling dimension of the field $\phi$ does not receive anomalous contributions.

We regularize $C^{(3)}_{\hat\sigma\hat\sigma}$ by adding a small
shift $\delta$ to the exponent of the internal line of the Hubbard-Stratonovich field $\hat\s$, obtaining
\begin{align}
 C^{(3)}_{\hat\sigma\hat\sigma}&{=}\frac{C_{\hat\sigma}\mu^{-\delta}}{|x|^{4{+}\delta}}4C_{\hat\phi}^2
 C_{\hat\sigma}\left(\frac{2\gamma_{\hat\phi}}{\delta}{+} A_{\hat\phi}\right)
 U\left(2,d{-}2{+}\frac{\delta}{2},{-}\frac{\delta}{2}\right)
U\left(2,\frac{d{+}\delta}{2},\frac{d{-}\delta}{2}{-}2\right)\,\frac{1}{(\mu |x|)^\delta}\notag\\
&=\frac{C_{\hat\sigma}}{|x|^4}\left(4\gamma_{\hat\phi}\log(\mu |x|)+\dots\right)\,,
\end{align}
where in the last line we took the limit $\delta\rightarrow 0$ and omitted everything except for the contribution to the anomalous dimension.
Consequently $C^{(3)}_{\hat\sigma\hat\sigma}$ contributes $-2\gamma_{\hat\phi}$ to
$\gamma_{\hat\sigma}$.
This contribution is precisely offset by the second term in the r.h.s. of (\ref{s->2 limit of gamma sigma}),
consistent with the fact that in the long-range CFT the counterpart diagram 
$C^{(3)}_{\sigma\sigma}$ is finite, and therefore does not provide any contributions to $\gamma_\sigma$.

Remarkably, our argument for continuity of scaling dimension of $\sigma$ does not use the specific value of
$\gamma_{\hat\phi}$. More importantly, this calculation pin-points how the
continuity of the scaling dimension of $\sigma$ is inherited from continuity of the scaling
dimension of $\phi$. Such an inheritance makes sense, since continuity of $\phi$ is intrinsically
connected with the irrelevance of the bi-local kinetic term for $\phi$ above the crossover
point. \\

Having established continuity of $\gamma_\sigma$, it is straightforward to generalize
our argument to demonstrate the continuity of $\gamma_{\sigma^2}$. We derived the latter
in section~\ref{sec:s2s2}.
The total dimension of $\sigma^2$ is given by $2s+\gamma_{\sigma^2}$, which at the crossover
point $s_\star = 2-2\gamma_{\hat\phi}$ can be rewritten as $4+(\gamma_{\sigma^2}|_{s\rightarrow2}-4\gamma_{\hat\phi})$.
Continuity at $s_\star$ requires this value to be equal to $4+\gamma_{\hat\s^2}$, the scaling  dimension
of $\hat\s^2$ in the short-range vector model. In other words, 
\begin{equation}
\label{s->2 limit of gamma sigma2}
\gamma_{\sigma^2}|_{s\rightarrow2}=\gamma_{\hat\sigma^2}+4\gamma_{\hat\phi}\,.
\end{equation}
One can see that almost all of the diagrams contributing to $\gamma_{\sigma^2}$
have a smooth limit $s\rightarrow 2$, under which they reduce to their short-range counterparts.
The only non-trivial diagram in this limit is the one with two $\langle\sigma\sigma\rangle$
sub-diagrams, where each of these receive $1/N$ corrections in the $s\rightarrow 2$ limit.
Then (\ref{s->2 limit of gamma sigma2}) immediately follows from (\ref{s->2 limit of gamma sigma}). It should be noted that since the anomalous dimensions of all operators $\s^n$ with $n>2$, come from the corrections to the $\s$ and $\s^2$ sub-diagrams, the continuity of their scaling dimensions across the crossover point follows from the continuity of the scaling dimensions of $\s$ and $\s^2$.  \\

To close this subsection, we will discuss continuity of
scaling dimension of the composite operator $\s\phi$.
Scaling dimension of this composite operator was
discussed in section~\ref{sec: sigma phi}, where we
provided a quick argument due to e.o.m. for the exact 
value of the anomalous dimension $\gamma_{\s\phi} = 0$
to all orders in $1/N$, as well as an explicit perturbative
calculation at the next-to-leading order in $1/N$ expansion.
Analogous calculation in the short-range vector model proceeds along the
similar steps, with setting $s\rightarrow 2$
in all the sub-leading diagrams. The only adjustment which one needs to make to the
long-range derivation is that the short-range $\hat\phi$ acquires anomalous dimension $\gamma_{\hat\phi}$
(the second and the third diagram in section~\ref{sec: sigma phi} contributing $\gamma_{\sigma\phi}^{(2,3)}$
have a continuous behavior at $s=2$).
Therefore the short-range counterpart of the first diagram in section~\ref{sec: sigma phi} scales as
\begin{align}
2(\Delta_{\hat\s}+\gamma_{\hat\s})+2(\Delta_{\hat\phi}+\gamma_{\hat\phi})
&=2(\Delta_\s|_{s\rightarrow 2}+\gamma_\s|_{s\rightarrow 2}-2\gamma_{\hat\phi})
+2(\Delta_{\phi}|_{s\rightarrow 2}+\gamma_{\hat\phi})\notag\\
&=\left[2(\Delta_\s+\gamma_\s)
+2\Delta_{\phi}\right]|_{s\rightarrow s_\star}
\end{align}
where we used (\ref{s->2 limit of gamma sigma}), consistently with the continuity at $s=s_\star = 2-2\gamma_{\hat\phi}$.

\subsection{Continuity of $\langle \phi\phi\sigma\rangle$ and $\langle \sigma\sigma\sigma\rangle$}

We now proceed to demonstrate continuity of normalized
three-point functions  $\langle \phi\phi\sigma\rangle$, $\langle \sigma\sigma\sigma\rangle$ at the
crossover point $s_\star$. Particularly, we are going to
demonstrate the continuity of the three-point function
amplitude (OPE coefficient) $\bar C_{\phi\phi\sigma}$
at the next-to-leading order in $1/N$ expansion,
as well as the amplitude $\bar C_{\sigma\sigma\sigma}$
at the leading order in $1/N$ expansion.

Recall that we split the amplitude into the leading large-$N$ factor
and the relative $1/N$ corrections as follows:
\begin{equation}
\label{Cphiphi s 1/N split}
\bar C_{\phi\phi\sigma}=\bar C_{\phi\phi\sigma}^{(0)}(1+\delta \bar C_{\phi\phi\sigma})\,.
\end{equation}
Here the leading order amplitude $\bar C_{\phi\phi\sigma}^{(0)}$
originates from the tree-level digram, while the relative sub-leading terms 
$\delta \bar C_{\phi\phi\sigma}$ are due to next-to-leading order corrections to the $\phi\phi \sigma$
vertex. The continuity requirement is then
\begin{align}
\label{continuity of Sphiphisigma}
\begin{split}
\bar C_{\hat\phi\hat\phi\hat\sigma}^{(0)}&=\bar C_{\phi\phi\sigma}^{(0)}\Bigg|_{s=2}\,,\\
\bar C_{\hat\phi\hat\phi\hat\sigma}^{(0)}\,\delta \bar C_{\hat\phi\hat\phi\hat\sigma} &=
\left(\bar C_{\phi\phi\sigma}^{(0)}\,\delta \bar C_{\phi\phi\sigma}
-2\gamma_{\hat\phi}\,\frac{\partial \bar C_{\phi\phi\sigma}^{(0)}}{\partial s}\right)\Bigg|_{s=2}\,.
\end{split}
\end{align}
Expressions for $\bar C_{\hat\phi\hat\phi\hat\sigma}^{(0)}$, $\delta \bar C_{\hat\phi\hat\phi\hat\sigma}$
are well-known \cite{Petkou:1994ad} and therefore 
the continuity (\ref{continuity of Sphiphisigma}) can be readily checked.
Using our result (\ref{hat Cphiphi sigma}), (\ref{delta C phiphi s in terms of Z})
we confirm that (\ref{continuity of Sphiphisigma}) is indeed satisfied.

Notice that while the first relation in (\ref{continuity of Sphiphisigma})
is a simple consequence of (\ref{s->2 simple limits}),
the meaning behind the second line in (\ref{continuity of Sphiphisigma}) is more opaque. 
However it does provide an interesting relation between the leading and next-to-leading
contributions to the three-point function $\langle\phi\phi\sigma\rangle$.\\

To close this section, we notice that
at the leading order in $1/N$, continuity of $\langle \sigma\sigma\sigma\rangle$ immediately follows
from  (\ref{continuity of Sphiphisigma}) and smooth behavior of the triangle diagram.


\section{A dual description for the long-range CFT at large $N$}
\label{sec: duality}

As we reviewed above, the long-range critical $O(N)$ vector model is defined
for the exponent $s$ taking values in the range $d/2< s< s_\star$.
It crosses over to the MFT regime for $s\leq d/2$, and to the short-range
 regime for $s\geq s_\star$ (given by critical vector model plus a decoupled
 generalized free field \cite{Behan:2017dwr,Behan:2017emf}). The model can be studied near the MFT transition point by setting $s=d/2+\epsilon$. This makes the quartic interaction slightly relevant, and allows one to study
the model perturbatively in  $\epsilon$ expansion. No such weakly coupled description of the short-range crossover point near $s=s_\star$ was available\footnote{With \cite{Sak1973} providing an earlier attempt of the construction 
perturbative near $s_\star$, as well as near $d=4$.} until a completely new model was suggested in  \cite{Behan:2017dwr} for the $N=1$ case (Ising model).

The proposal was
 to start with the action $S_{\textrm{crit}}$ for the short-range critical Ising model ($N=1$ vector model), and
 couple it to a generalized free field $\chi$,
\begin{align}
 S = S_{\textrm{crit}} +\int d^dx\int d^dy \,\frac{\chi(x)\chi(y)}{|x-y|^{d-s}}+ \l\int\,d^dx \,\hp\chi \,.
\end{align}
The scaling dimension of $\chi$ is therefore fixed to be $\Delta_{\chi}=\f{d+s}2$,
while the field $\hat\phi$ (which is the lowest scalar primary of the critical Ising CFT)
has dimension $\frac{d}{2}-1+\gamma_{\hat\phi} =  \f{d-s_\star}2$, when $\lambda = 0$.
%
The dimension of $\chi$ was tuned such that the perturbation $\hp\chi$ is irrelevant for $s>s_\star$, and relevant for $s<s_\star$.
In the latter case, the coupling $\lambda$
triggers an RG flow that
can be studied perturbatively in the vicinity of $s_\star$.
Interestingly, the IR fixed point of this flow is described by a long-range CFT.
Moreover, it has been suggested  in \cite{Behan:2017dwr} that the resulting long-range CFT is,
for all $d/2\leq s\leq s_\star$,
the same as the long-range critical vector model, obtained at the end of the RG flow of the Gaussian MFT deformed by a quartic operator. Therefore, these two UV descriptions are dual in the IR,
and have complementary perturbative regimes on the opposite ends of the range $d/2\leq s\leq s_\star$. However, evidence for the duality for all intermediate values of $s$ has so far been obtained only for certain three-point function coefficient ratios, valid for all orders in the $\epsilon$-expansion \cite{Behan:2017dwr}.

In this section, we aim to provide a complete non-perturbative description of the duality for all values of $s$, by reformulating the duality suggested in  \cite{Behan:2017dwr} at large $N$. We begin with re-writing the action for the short-range critical $O(N)$ vector model at large $N$
in the Hubbard-Stratonovich formalism,
\begin{align}
\label{critical action of On vm}
 S_{\textrm{crit}} = \int\,d^dx \left(\f12(\p_{\mu}\hp)^2-\f{\hs^2}{4g} + \f1{\sqrt{N}}\hs\hp^2\right)+\cdots\,,
\end{align}
where ellipsis stands for $1/N$ corrections, and we also skip explicitly writing the counterterms.
Following \cite{Behan:2017dwr}, we introduce a new field $\chi^i$, $i=1,\dots,N$
with a bi-local kinetic term, and scaling dimension
\begin{equation}
\Delta_{\chi} = \frac{d+s}{2}\,,
\end{equation}
and couple it to the field $\hat\phi^i$ as follows:
\begin{equation}
\label{original coupling lambda}
S =  S_{\textrm{crit}} + a\,\int d^dx\int d^dy\,\frac{\chi^i(x) \chi^i(y)}{|x-y|^{d-s}} + \lambda\,
\int d^dx\,\hat\phi^i\chi^i\,.
\end{equation}
Here, the scaling dimension of $\hat\phi^i$ is $(d-s_\star)/2$.
The kinetic term normalization coefficient $a$ can be arbitrary, and
reflects conventions regarding the definition of $\chi$. A convenient choice to fix it will be explained momentarily.
To obtain a large $N$ description of the IR CFT, we first perform the following redefinition of the fields and couplings,
\begin{equation}
\label{Phi Sigma def}
\Phi^i = \lambda\hat\phi^i\,,\qquad \Sigma = \frac{1}{\lambda^2}\hat\sigma\,,\qquad g= \lambda^4 G \,.
\end{equation}
This allows us to re-write the action as\footnote{We skip keeping track of the $O(N)$ indices.}
\begin{equation}
\label{chi coupled to sr model}
S_{\textrm{SR}\chi} = {a}\int d^dx\int d^dy\,\frac{\chi(x) \chi(y)}{|x{-}y|^{d{-}s}} 
{+}\int d^dx\,\left( \frac{1}{2\lambda^2}\,(\partial\Phi)^2
{-}\frac{1}{4G}\Sigma^2 {+}\Phi\chi{+}\frac{1}{\sqrt{N}}\Sigma\Phi^2 \right)+\cdots\,,
\end{equation}
where we have again omitted higher-order corrections to vertices and propagators.

We will now argue that when both $G$ and $\lambda$ are tuned to criticality,
and $s$ assumes values in the range $d/2<s<s_\star$,
the model (\ref{chi coupled to sr model}) is equivalent to the long-range CFT (\ref{eff-action-1}).
The argument can be constructed by going from the latter towards the former. 

Starting from the long-range CFT model (\ref{eff-action-1}),
we can add to it its action a local kinetic term $\frac{1}{2\lambda^2}\,(\partial\phi)^2$,
where $\lambda$ is a dimensionful constant that we choose to be equal to the IR critical
value of the coupling constant in the model (\ref{chi coupled to sr model}).
This can be done without affecting dynamics of the field $\phi$ in the range $d/2<s<s_\star$,
where such a term is suppressed compared to the bi-local kinetic term for the field $\phi$.

We then insert an identity
into the partition function $Z_{\textrm{LR}}$ of the model (\ref{eff-action-1}),
represented as a Gaussian integral over a field $\chi$\footnote{Our notation for this
field is deliberately the same as for the $\chi$ field in the action (\ref{chi coupled to sr model}),
as we intend to identify these d.o.f. on both sides of the argued IR duality.}
\begin{align}
Z_{\textrm{LR}}&\rightarrow Z_{\textrm{LR}}\,\frac{1}{z}\,
\int [D\chi]\exp\left(-a\int \frac{d^dx_{1,2}}{|x_{12}|^{d-s}}
\left(\chi+\alpha\int d^dx_1'\,\frac{\phi(x_1^\prime)}{|x_1-x_1^\prime|^{d+s}}\right)\right.\notag\\
&\times\left.\left(\chi+\alpha\int d^dx_2^\prime \,\frac{\phi(x_2^\prime)}{|x_2-x_2^\prime|^{d+s}}\right)\right)\,,
\end{align}
accompanied by the corresponding normalization prefactor $1/z$.
Choosing
\begin{align}
a &= -\frac{1}{4C(s)\pi^d A(\Delta_{\Phi})A(d-\Delta_{\Phi})}\,,\\
\alpha &=\frac{1}{2a\pi^d A(\Delta_{\Phi})A(d-\Delta_{\Phi})} = -2C(s)\,,
\end{align}
we obtain that the action (\ref{eff-action-1}) transforms into
\begin{align}
\label{HS transformed LR action}
S_{\textrm{LR}}=  a\int d^dx\int d^dy\,\frac{\chi(x) \chi(y)}{|x-y|^{d-s}} 
+ \int\,d^dx \left(\frac{1}{2\lambda^2}\,(\partial\phi)^2-\f1{4g} \s^2+\phi\chi + \f1{\sqrt{N }} \s \phi^2\right)\,.
\end{align}
Under the identification 
\begin{equation}
\label{hat Phi and hat Sigma correspondence}
\Phi\leftrightarrow \phi\,,\qquad \Sigma\leftrightarrow\sigma\,,\qquad g\leftrightarrow G
\end{equation}
the models (\ref{chi coupled to sr model}), (\ref{HS transformed LR action}) are indeed equivalent.

Notice that when $s>s_\star$ the coupling $\lambda$ in  (\ref{original coupling lambda})
becomes irrelevant, and its critical value is a trivial $\lambda = 0$, making the field
redefinition (\ref{Phi Sigma def}) ill-defined. In fact, the field $\chi$ decouples into
an independent d.o.f. with a bi-local kinetic term.
Therefore while the IR duality between the long-range vector model
and the deformed short-range critical vector model, defined for $d/2<s<s_\star$,
is quite straightfoward to establish, a more subtle prediction is that when $s>s_\star$
(outside the scope of the long-range model, and therefore beyond the regime
where one can talk about such a duality) the system retains a decoupled
generalized free field d.o.f \cite{Behan:2017dwr,Behan:2017emf}.\footnote{The dual 
description of the critical long-range model is also useful near $s=s_\star$, where it
gives a perturbative handle on the calculations \cite{Behan:2017dwr,Behan:2017emf}.}

Having argued for the duality between the long-range and the deformed short-range
critical vector models in the range $d/2<s<s_\star$, we would like to
explore the meaning of the dual 
of the generalized free field $\chi$ on the long-range side, with an aim to make the duality more precise.
Motivated by this goal, for the rest of this section we are going to return to the picture with the original field $\chi$. The resulting action
of the critical regime of the $O(N)$ vector model coupled to $\chi$
is given by
\begin{equation}
\label{chi coupled to sr model short}
S = -\frac{1}{4C(s)\pi^d A(\Delta_{\Phi})A(d-\Delta_{\Phi})}\,\int d^dx\int d^dy\,\frac{\chi(x) \chi(y)}{|x-y|^{d-s}} 
+\int d^dx\,\left( \Phi\chi+\frac{1}{\sqrt{N}}\Sigma\Phi^2 \right)\,.
\end{equation}
\noindent
It follows from the action (\ref{chi coupled to sr model short}) that the propagator of the field $\chi$ is
\begin{equation}
\langle \chi(x)\chi(0)\rangle = \frac{C_\chi\,(1+A_{\chi})}{|x|^{2\Delta_{\chi}}}\,,\qquad C_\chi = -2C(s)\,,
\end{equation}
where we took into account that the dimension $\Delta_{\chi}$ is exact, and the only 
possible $1/N$ corrections to the $\chi$ propagator can go into the amplitude 
correction $A_{\chi}$.
At the same time, the $\chi$ e.o.m. following from (\ref{chi coupled to sr model short}) is,
\begin{equation}
\label{Phi chi relation due to eom}
\Phi = \frac{1}{2C(s)\pi^d A(\Delta_{\Phi})A(d-\Delta_{\Phi})}\int d^dy\,\frac{\chi(y)}{|x-y|^{d-s}}\,,
\end{equation}
which, in turn, gives
\begin{equation}
\langle\Phi(x)\Phi(0)\rangle = 
-\frac{1+A_{\chi}}{2\pi^dC(s)A(\Delta_{\Phi})A(d-\Delta_{\Phi})}\,\frac{1}{|x|^{2\Delta_{\Phi}}}\,.
\end{equation}

Important manifestation of a duality between two CFTs is given by matching
conformal correlation functions on both sides of the duality.
When matching amplitudes of the
 correlation functions of operators in different models, one needs to ensure consistent normalization of these
 operators. A convenient choice is given by normalization of the two-point functions
 of the considered operators to unity.\footnote{See \cite{Goykhman:2019kcj,Goykhman:2020tsk,Chai:2021uhv} for a recent discussion.}
Particularly, one finds it useful that rescaling the fields $\chi$, $\Phi$ as
\begin{align}
\label{normalization of hat chi and hat Phi}
\chi\rightarrow\frac{1}{\sqrt{-2C(s)\,(1+A_{\chi})}}\,\chi\,,\quad
\Phi\rightarrow\sqrt{-\frac{2\pi^dC(s)A(\Delta_{\Phi})A(d-\Delta_{\Phi})}{1+A_{\chi}}}\,\Phi\,,
\end{align}
obtaining unit-normalized propagators $\langle\chi\chi\rangle$,
$\langle\Phi\Phi\rangle$. The three-point functions involving the fields $\chi$, $\Phi$
will be calculated for such fields normalized according to (\ref{normalization of hat chi and hat Phi}).
Similar normalization conventions can be used for
 their counterparts in the long-range model.
 
 In particular, one can establish that the ratios of coefficients of three-point
 functions $\langle \chi O_2O_3\rangle$, $\langle \Phi O_2O_3\rangle$,
 and $\langle \sigma\phi O_2O_3\rangle$, $\langle \phi O_2O_3\rangle$
where $O_{2,3}$ are some conformal operators, match exactly to all orders in $1/N$
and for all $d/2<s<s_\star$, analogously to \cite{Behan:2017dwr,Behan:2017emf}.
 This implies the following matching of d.o.f. on both sides of the duality:
 \begin{equation}
\label{chi and sigmaphi relation}
\chi\leftrightarrow\sigma\phi\,.
\end{equation}
Simply put,
the relation (\ref{chi and sigmaphi relation}) between $\chi$ and $\sigma\phi$
follows from the relation (\ref{hat Phi and hat Sigma correspondence}) $\Phi\leftrightarrow \phi$, accompanied by the observation
that the pairs of fields  $\sigma\phi, \phi$ and $\chi, \Phi$  satisfy analogous equations of motion
(\ref{phi eom in LR}), (\ref{Phi chi relation due to eom}).\\

An infra-red duality between the critical long-range
model and the short-range model deformed by a coupling to the generalized free field $\chi$,
requires one to match the spectrum of the two CFTs,
as well as all of the CFT data. An immediate observation one can make is that
in the $N=1$ case the long-range model possesses an operator $\phi^3$
with exact scaling dimension $(d+s)/2$,  while the short-range model has no analogous
counterpart \cite{Behan:2017dwr}. The corresponding operator in
the critical long-range $O(N)$ model in the 
Hubbard-Stratonovich language is $\sigma\phi$,
whose dimension $\Delta_{\s\phi}$
given by (\ref{Delta sigma phi}), as we derived in section~\ref{sec: sigma phi}, is fixed exactly by the e.o.m. relating it to the dimension of $\phi$.
Together with the other operator $\phi$
of a fixed scaling dimension (\ref{Delta phi}), the `shadow relation' is satisfied,
analogously to the $N=1$ case of \cite{Paulos:2015jfa}
\begin{equation}
\Delta_\phi + \Delta_{\sigma\phi} =d\,.
\end{equation}

Notice that absence of the anomalous dimension contribution to the composite operator $\s\phi$
in the long-range critical model is to be contrasted with its counterpart $\hat\s\hat\phi$ in the short-range
vector model. This being said, the full scaling dimension of $\s\phi$ is continuous across $s_\star$,
as we demonstrated in section~\ref{sec: continuity of dimensions}.\\

We close this subsection by pointing out that
using e.o.m. (\ref{phi eom in LR}) we can also calculate the following cross-correlator
\begin{equation}
\label{<sigma phi phi>}
\langle \sigma\phi(x)\phi(0)\rangle = -\frac{C(s)}{\sqrt{N}}\,C_\phi\,
\pi^d A\left(\frac{d-s}{2}\right)A\left(\frac{d+s}{2}\right)\,\delta^{(d)}(x)\,,
\end{equation}
which vanishes for non-coincident points. The counterpart of this correlator
in the deformed short-range model is given by $\langle \chi(x)\hat\phi(0)\rangle\simeq \delta^{(d)}(x)$.
Notice that the result (\ref{<sigma phi phi>}) is exact to all orders in $1/N$, but can
be seen explicitly at the leading order in $1/N$ from the diagram
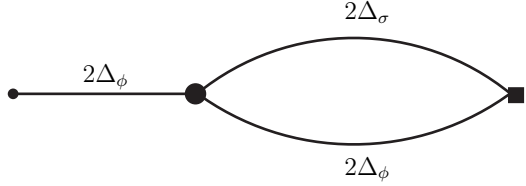
\begin{center}
  \begin{picture}(368,65) (-70,10)
    \SetWidth{1.0}
    \SetColor{Black}
    \Vertex(-20,38){2}
    \Vertex(48,38){4}
    \Line[](-20,38)(48,38)
    \Arc[clock](107,-32.976)(91.978,129.496,50.504)
    \Arc[](107,117.5)(98.501,-127.161,-52.839)
    \CBox(165,35)(170,40){Black}{Black}
    \Text(104,65)[lb]{\scalebox{0.8}{$2\Delta_\s$}}
    \Text(104,5)[lb]{\scalebox{0.8}{$2\Delta_\phi$}}
    \Text(6,40)[lb]{\scalebox{0.8}{$2\Delta_\phi$}}
  \end{picture}
\end{center}

\subsection{Anomalous dimension of the stress-energy tensor}

In this section, we continue studying the $\hat\phi\chi$ perturbation (\ref{original coupling lambda})
of the short-range $O(N)$ vector model (\ref{critical action of On vm}) that brings it to a long-range
critical regime in the IR. Our focus will be on the fate of the
stress-energy tensor operator $T_{\mu\nu}$,
\begin{equation}
\label{Tmunu}
T_{\mu\nu} = \partial_\mu\hat\phi\partial_\nu\hat\phi 
-\delta_{\mu\nu}\,\left(\frac{1}{2}\,(\partial\hat\phi)^2-\frac{1}{4g_\star}\,\hat\sigma^2+
\frac{1}{\sqrt{N}}\,\hat\sigma\hat\phi^2\right)\,
\end{equation}
of the original short-range vector model (\ref{critical action of On vm})
under such a deformation.
In (\ref{Tmunu}) we have substituted the critical value of the coupling $g_\star$.
One anticipates that conservation of the composite operator (\ref{Tmunu})
is broken at the long-range fixed point of the model (\ref{original coupling lambda})  \cite{Behan:2017dwr,Behan:2017emf},
where it no longer has the status of a conserved stress-energy tensor.
In agreement with such an expectation, working at the first order in $1/N$ expansion, we will derive
anomalous dimension and trace of the operator (\ref{Tmunu}). 

Unlike calculation of the previous section (that was carried out in the entire range $d/2<d<s_\star$),
in this section we will stay perturbatively close to the cross-over point $s_\star$,
working at the linear order in $\delta =(s_\star - s)/2$. Since our starting point
is the short-range critical vector model, we find it convenient for the purposes
of this section to work in terms of the original conventions $\hat\phi$, $\hat\sigma$
for the degrees of freedom of this model.

While in the original theory, $\lambda=0$, the stress-energy tensor (\ref{Tmunu})
is conserved on shell,  conservation law of that operator is broken at the new fixed point,
where it no longer plays the role of stress-energy tensor.
This can manifest in the anomalous dimension $\gamma_T$ and trace $\hat c$,
in the ansatz for the correlation function \footnote{Here we skipped
$2\gamma_T$ in the exponent in the second term in the r.h.s., anticipating linearization
in $\delta$, while taking into account that both $\gamma_T$ and $\hat c$ vanish when $\delta =0$.}
\begin{align}
\label{TT general}
 \lan \partial_\mu T^{\mu\nu}(x_1)\partial_\lambda T^{\l\rho}(x_2)\ran =\frac{\partial}{\partial x_1^\mu}
 \frac{\partial}{\partial x_2^\lambda}\left( C_T \f{I^{\mu\nu,\l\rho}(x_{12})}{|x_{12}|^{2d+2\gamma_T}}
 +\hat c\,\delta^{\mu\nu}\,\delta^{\lambda\rho}\,\frac{1}{|x_{12}|^{2d}}\right)\,,
\end{align}
where $\gamma_T$ is the anomalous dimension, 
$C_{T}$ is the central change of the short-range theory,
\begin{equation}
C_T = N\,\left(\frac{d}{(d-1)S_d^2}+{\cal O}\left(\frac{1}{N}\right)\right)\,,\qquad
S_d = \frac{2\pi^\frac{d}{2}}{\Gamma\left(\frac{d}{2}\right)}\,,
\end{equation}
We also defined tensor structures 
\begin{align}
I_{\mu\nu}(x) &= \delta_{\mu\nu}-2\f{x_\mu x_\nu}{x^2}\,,\\
 I_{\mu\nu,\l\rho}(x) &= \f12\left(I_{\mu\l}(x)I_{\nu\rho}(x) + I_{\mu\rho}(x)I_{\nu\l}(x)-\f2{d}\delta_{\mu\nu}\delta_{\l\rho}\right)\,,
\end{align}
and the $\hat c$ term in (\ref{TT general}) accounts for a possible non-vanishing trace.\footnote{We thank M.~Smolkin for discussion of this point.}
Simplifying (\ref{TT general}), while linearizing in $\delta$, we obtain
\begin{equation}
\label{del T del T general}
 \lan \partial_\mu T^{\mu\nu}(x)\partial_\l T^{\l\rho}(0)\ran
= \frac{C_T(d-1)(d+2)\gamma_T}{d}\,
I^{\nu\rho}\,\frac{1}{|x|^{2d+2}}+2d\hat c \,\left(I^{\nu\rho}-2d\frac{x^\nu x^\rho}{|x|^2}\right)\,\frac{1}{|x|^{2d+2}}\,.
\end{equation}

We are going to compare the general ansatz (\ref{del T del T general}) with what we obtain
specifically in the model (\ref{original coupling lambda}). Working at the second order in conformal
perturbation theory in $\lambda_\star$ we derive
\begin{align}
 &\lan \partial_\mu T^{\mu\nu}(x)\partial_\l T^{\l\rho}(0)\ran =\frac{\lambda_\star^2}{2}\,\int d^dx_{1,2}\,
 \langle  \partial_\mu T^{\mu\nu}(x)\partial_\l T^{\l\rho}(0) \hat\phi(x_1)\chi(x_1)
 \hat\phi(x_2)\chi(x_2) \rangle +{\cal O}(\lambda_\star^3)\notag\\
 &=\frac{\lambda_\star^2}{2}\,\int d^dx_{1,2}\,\langle \chi (x_1)\chi(x_2) \rangle\,
 \langle \partial_\mu T^{\mu\nu}(x)\partial_\l T^{\l\rho}(0) \hat\phi(x_1) \hat\phi(x_2) \rangle +{\cal O}(\lambda_\star^3)\notag\\
  &=\lambda_\star^2\,\langle \chi (x)\chi(0) \rangle\,\langle\partial ^\nu\hat\phi(x)\partial^\rho \hat\phi(0) \rangle 
  +{\cal O}(\lambda_\star^3)\,,
  \label{del T del T Ward}
\end{align}
where in the last line we used Ward identity for the divergence of the stress-energy tensor in the
original short-range sector of the model.
From (\ref{del T del T Ward}) we then obtain
\begin{equation}
\label{del T del T on shell}
\langle \partial_\mu T^{\mu\nu}(x)\partial_\lambda T^{\lambda\rho}(0)\rangle
=N\lambda_\star ^2 (d-s_\star)C_{\hat\phi} C_\chi \,\frac{1}{|x|^{2d+2}}\,\left(I^{\nu\rho}
-(d-s_\star)\,\frac{x^\nu x^\rho }{|x|^2}\right)\,,
\end{equation}
where we substituted $s=s_\star$ at the short-range fixed point.

Comparing (\ref{del T del T on shell}), (\ref{del T del T general}) while demanding that tensor
structures match, we arrive at
\begin{align}
\hat c &=N\lambda_\star^2 \frac{(d-s_\star)^2 C_{\hat\phi} C_\chi}{4d^2}+{\cal O}(\lambda_\star^3)\,,\label{hat c result}\\
\gamma_T &= N\lambda_\star^2 \frac{(d+s_\star)(d-s_\star) C_{\hat\phi} C_\chi}{2C_T (d-1)(d+2)}+{\cal O}(\lambda_\star^3)\,.
\label{gamma T result}
\end{align}
In the perturbative regime, near the short-range fixed point, we have \cite{Behan:2017dwr}
\begin{equation}
\lambda_\star^2 = \alpha (d)\delta +{\cal O}(\delta^2)\,,
\end{equation}
where $\delta =(s_\star-s)/2$. Furthermore, plugging $s_\star=2-2\gamma_{\hat\phi}$
in (\ref{hat c result}), (\ref{gamma T result}) and expanding
in $\gamma_{\hat\phi}$ (which in the large $N$ language means
expanding in $1/N$), we obtain 
\begin{align}
\hat c &=N\,\alpha\,\delta\, \frac{(d-2)\Gamma\left(\frac{d}{2}+1\right)^2}
{d^3 \pi^d}\,\gamma_{\hat\phi}+{\cal O}(\delta ^2,\gamma_{\hat\phi}^2)\,,\label{hat c result 2}\\
\gamma_T &= \frac{2\,\alpha\,\delta}{\Gamma\left(\frac{d}{2}\right)^2}\,
\gamma_{\hat\phi}+{\cal O}(\delta ^2,\gamma_{\hat\phi}^2)\,.
\label{gamma T result 2}
\end{align}

While the anomalous dimension and trace (\ref{hat c result 2}), (\ref{gamma T result 2})
have been derived perturbatively close to the short-range fixed point,
the operator $\partial_\mu T^{\mu\nu}$ can be studied in the entire range $d/2<s<s_\star$.
We can  decompose this operator in terms of the primary  \cite{Behan:2017dwr,Giombi:2016hkj}
\begin{equation}
\label{Vnu definition}
V^\nu = \hat\phi\partial^\nu\chi - \frac{\Delta_\chi}{\Delta_\phi}\,\chi\partial^\nu\hat\phi\,,
\end{equation}
and the descendant $\partial^\nu (\hat\phi\chi)$. Notice that the latter has a vanishing (leading order)
cross-correlator with (\ref{Vnu definition}). To find coefficients of such a decomposition, we will be
working in the conformal perturbation theory at linear order in $\lambda_\star$:
\begin{align}
\langle \del _\mu T^{\mu\nu}(x) \partial^\rho (\hat\phi\chi)(0)\rangle
&=\lambda_\star\,\int d^dx_1\,\langle \del _\mu T^{\mu\nu}(x)\, \partial^\rho (\hat\phi\chi)(0)\,\hat\phi\chi(x_1)\rangle\notag\\
&=\lambda_\star \langle\partial^\nu\hat\phi\, \chi(x) \, \partial^\rho(\hat\phi\chi)(0)\rangle\,,
\end{align}
where in the last line we used Ward identity. Similarly, we obtain
\begin{align}
\langle \del _\mu T^{\mu\nu}(x) V^\rho(0)\rangle
=\lambda_\star \langle\partial^\nu\hat\phi\, \chi(x) \, V^\rho(0)\rangle\,.
\end{align}
Therefore
\begin{equation}
\del _\mu T^{\mu\nu}(x)  = \lambda_\star\partial^\nu\hat\phi\, \chi + {\cal O}(\lambda_\star^2)\,,
\end{equation}
which we can rewrite as 
\begin{equation}
\label{partial mu Tmunu decomposition}
\partial_\mu T^{\mu\nu} = -\lambda_\star\,\frac{d-s}{2d}\,V^\nu - \frac{d-s}{2d}\,\lambda_\star\,\partial^\nu (\Phi\chi)\,.
\end{equation}

A short-cut to the calculation of $\gamma_T$, providing
a faster alternative derivation of the result (\ref{gamma T result}),
is afforded by the decomposition (\ref{partial mu Tmunu decomposition}).
One can easily see that focusing only on the $V^\nu$
contribution to the divergence of the stress-energy tensor (\ref{partial mu Tmunu decomposition}),
while simultaneously ignoring that trace contribution $\hat c$ in 
the general ansatz (\ref{del T del T general}), reproduces the final expression $(\ref{gamma T result})$
for the anomalous dimension $\gamma_T$ \cite{Behan:2017dwr,Behan:2017emf}.\footnote{In particular,
one can see that the coefficient in front of the $V^\nu$ term
in r.h.s. of (\ref{partial mu Tmunu decomposition}) matches with its counterpart
$b(g_\star)=b_1g_\star+{\cal O}(g_\star^2)$ introduced in \cite{Behan:2017dwr,Behan:2017emf}.}
Indeed, following such a strategy we obtain from (\ref{partial mu Tmunu decomposition})
\begin{equation}
\label{del T del T with just VV}
\lan \partial_\mu T^{\mu\nu}(x)\partial_\lambda T^{\l\rho}(0)\ran
=\lambda_\star^2\,\frac{(d-s)^2}{4d^2}\,\lan V^\nu (x) V^\rho (0)\ran +\dots \,.
\end{equation}
Substituting here
\begin{equation}
\langle V^\nu(x) V^\rho(0)\rangle = N\,\frac{2dC_{\hat\phi}C_\chi (d+s)}{d-s}\,\frac{1}{|x|^{2d+2}}\,I^{\nu\rho}\,,
\end{equation}
and comparing the result with the $\gamma_T$ term in (\ref{del T del T general})
we recover (\ref{gamma T result}).

\section{Discussion}
\label{sec:discussion}

The main focus of this paper was to study the long-range critical $O(N)$ vector model
in the large-$N$ limit. This model appears in the IR regime of the Gaussian MFT for a generalized free field, deformed by a local quartic interaction.
Working within the Hubbard-Stratonovich formalism, we calculated several new
scaling dimensions and OPE coefficients of various primary operators in this model,
performing most of our calculations at the next-to-leading order in the $1/N$ expansion.
The CFT data we obtained furnishes a non-trivial consistency check
for the existence of the full conformal symmetry at the long-range fixed point.

In particular, we determined the leading order contribution to the anomalous dimension
of the composite operator $\sigma^n$, $n\geq 2$, and calculated the three-point
functions $\langle \phi\phi\s\rangle$, $\langle \s\s\s\rangle$. We also established that the cross-correlator
$\langle \s^2\s\rangle$ vanishes, at least at the leading order in $1/N$, unlike its short-range counterpart discussed in
\cite{Derkachov:1997gc,Derkachov:1998js}. The vanishing of this correlator simultaneously indicates that the long-range fixed point is indeed a CFT, and that $\s^2$ is in fact a primary operator in this CFT.

While in the short-range CFT certain composite operators, such as $\hat\s^2$, can mix with
the descendants, such as $\partial^2\hat\s$, such a mixing is impossible in the long-range CFT for the
Hubbard-Stratonovich field with a leading-order scaling dimension $s$, taking a general value in the 
range $d/2<s<s_\star$ (see \cite{Behan:2018hfx} for a bootstrap analysis of the long-range Ising model).
The case of $s=1$ requires a special attention, since it allows for the possibility of operator
mixing, at least in principle, such as a mixing between $\s^4$ and $\partial^2\s^2$. Notice that $s=1$
in fact has the physical interpretation of a boundary CFT (BCFT),
\textit{i.e.}, a free field theory in the $d+1$-dimensional bulk perturbed by a 
quartic interaction localized on its $d$-dimensional boundary (or defect).\footnote{For the quartic
operator to be relevant in the IR one needs to consider $d<2$. When $d>2$ the theory arrives
at the critical regime in the UV, however it suffers from usual instabilities typically found in such cases \cite{Giombi:2019enr}.}
We therefore expect the
spectrum of conformal primaries to shrink in the BCFT, as contrasted to a generic
long-range CFT. It would be interesting to explore this direction further.

We have also investigated the interplay between long-range and short-range critical vector models.
While the former is expected to occupy the region $d/2<s<s_\star$ of the exponent parameter $s$,
the latter exists in the region $s>s_\star$. We have performed an explicit consistency check
of our results for the calculated CFT data, by demonstrating its continuity at the
long-range--short-range crossover point $s_\star$. 
Furthermore, generalizing the construction of \cite{Behan:2017dwr,Behan:2017emf} to the large-$N$
case, we have argued for the existence of an exact IR duality between the long-range vector
model, and the short-range model deformed by the coupling to a generalized free field.
Performing a simultaneous
perturbative expansion to the linear order in $s_\star-s$ and to the next-to-leading order in $1/N$
we also obtained expression for the anomalous dimension and trace of the stress-energy tensor
of the short-range vector model, which it acquires at the long-range fixed point.

To close this discussion, we would like to notice that long-range critical vector models
has recently received attention from the perspective of persistent symmetry breaking
at all temperatures \cite{Chai:2021djc}, adding extra motivation for expanding our understanding
of CFT data of these models.

\section*{Acknowledgements} \noindent  We thank M.~Smolkin, S.~Rychkov for helpful discussions. Our work is partially supported by the Binational Science Foundation (grant No. 2016186), the Israeli Science Foundation Center of Excellence (grant No. 2289/18), and by the Quantum Universe I-CORE program of the Israel Planning and Budgeting Committee (grant No. 1937/12). The work of NC is partially supported by Yuri Milner scholarship.

\appendix

\section{Some useful identities}
\label{app_a}

In this appendix we collect some useful expressions and identities.

Loop diagram in the position space are simply additive:
\begin{center}
  \begin{picture}(257,50) (0,0)
    \SetWidth{1.0}
    \SetColor{Black}
    \Arc[clock](81,-39)(77.006,127.614,52.386)
    \Arc[](80,86)(80,-126.87,-53.13)
    \Line[](160,22)(256,22)
    \Vertex(160,22){2}
    \Vertex(256,22){2}
    \Vertex(33,22){2}
    \Vertex(129,22){2}
    \Text(142,20)[lb]{$=$}
    \Text(77,-5)[lb]{\scalebox{0.8}{$2b$}}
    \Text(77,43)[lb]{\scalebox{0.8}{$2a$}}
    \Text(190,27)[lb]{\scalebox{0.8}{$2(a+b)$}}
  \end{picture}
\end{center}

The propagator merging relation is given by
\begin{equation}
\int d^d x_2\, \frac{1}{|x_2|^{2a}|x_1-x_2|^{2b}}
=U(a,b,d-a-b)\,\frac{1}{|x_{1}|^{2a+2b-d}}\,,
\end{equation}
where we introduced
\begin{align}
U(a,b,c) &= \pi^\frac{d}{2} A(a)A(b)A(c)\,.
\end{align}
This relation can be diagrammatically represented as
\begin{center}
  \begin{picture}(98,10) (130,-60)
    \SetWidth{1.0}
    \SetColor{Black}
    \Line[](30,-58)(90,-58)
    \Line[](90,-58)(150,-58)
    \Line[](180,-58)(240,-58)
    \Vertex(30,-58){2.001}
    \Vertex(90,-58){4.001}
    \Vertex(150,-58){2.001}
    \Vertex(180,-58){2.001}
    \Vertex(240,-58){2.001}
    \Text(55,-53)[lb]{\scalebox{0.81}{$2a$}}
    \Text(115,-53)[lb]{\scalebox{0.81}{$2b$}}
    \Text(163,-61)[lb]{$=$}
    \Text(180,-53)[lb]{\scalebox{0.81}{$2(a+b)-d$}}
    \Text(250,-63)[lb]{$\times U(a,b,d-a-b)$}
  \end{picture}
\end{center}

Uniqueness relation for $a_1+a_2+a_3=d$ is written as \cite{DEramo:1971hnd,Symanzik:1972wj}
\begin{equation}
\int d^dx\,\frac{1}{|x_1-x|^{2a_1}|x_2-x|^{2a_2}|x_3-x|^{2a_3}}
=\frac{U(a_1,a_2,a_3)}{|x_{12}|^{d-2a_3}|x_{13}|^{d-2a_2}|x_{23}|^{d-2a_1}}\,,
\end{equation}
and is graphically represented as
\begin{center}
  \begin{picture}(210,66) (70,-31)
    \SetWidth{1.0}
    \SetColor{Black}
    \Line[](32,34)(80,2)
    \Line[](80,2)(32,-30)
    \Line[](80,2)(128,2)
    \Line[](192,34)(192,-30)
    \Line[](192,-30)(240,2)
    \Line[](240,2)(192,34)
    \Vertex(32,34){2.01}
    \Vertex(32,-30){2.01}
    \Vertex(80,2){4.01}
    \Vertex(128,2){2.01}
    \Vertex(192,34){2.01}
    \Vertex(192,-30){2.01}
    \Vertex(240,2){2.01}
    \Text(55,-25)[lb]{\scalebox{0.81}{$2a_1$}}
    \Text(55,22)[lb]{\scalebox{0.81}{$2a_2$}}
    \Text(100,5)[lb]{\scalebox{0.81}{$2a_3$}}
    \Text(160,-1)[lb]{$=$}
    \Text(180,-1)[lb]{\scalebox{0.81}{$\alpha$}}
    \Text(215,-28)[lb]{\scalebox{0.81}{$\beta$}}
    \Text(215,24)[lb]{\scalebox{0.81}{$\gamma$}}
    \Text(250,-5)[lb]{$\times \left(-\frac{2}{\sqrt{N}}\right) U\left(a_1,a_2,a_3\right)$}
  \end{picture}
\end{center}
Here we defined $\alpha = d-2a_3$, $\beta = d-2a_2$, $\gamma = d-2a_1$.

\newpage
\bibliographystyle{utphys}
\providecommand{\href}[2]{#2}\begingroup\raggedright\endgroup

\end{document}